\documentclass[12pt,a4paper,fleqn]{article}
\usepackage[utf8]{inputenc}
\usepackage{authblk}
\usepackage{geometry}
\usepackage{xcolor}
\usepackage[round]{natbib}
\usepackage{graphicx}
\usepackage{pdflscape}
\usepackage{lscape}
\usepackage{appendix}
\usepackage{setspace}
\usepackage{dashrule}
\usepackage[shortlabels]{enumitem}
\usepackage{longtable}
\usepackage{chngcntr}
\usepackage{afterpage}

\usepackage{colortbl}

\providecommand{\shadeRow}{\rowcolor[rgb]{0.9, 0.9, 0.9}}
\providecommand{\shadeBench}{\rowcolor[rgb]{0.95, 0.3, 0.3}}

\usepackage{threeparttable}
\usepackage{booktabs}

\usepackage{amsmath}
\usepackage{amsfonts}

\usepackage{bm}

\usepackage{enumitem}
\newlist{steps}{enumerate}{1}
\setlist[steps, 1]{label = Step \arabic*:}

\usepackage{dcolumn}
\newcolumntype{d}[1]{D{.}{.}{#1}}
\usepackage{soul}


\usepackage{setspace}

\usepackage{caption}
\usepackage{subcaption}
\definecolor{nblue}{HTML}{000660}
\usepackage[colorlinks=true,urlcolor=nblue,linkcolor=nblue,citecolor=nblue]{hyperref}

\title{\Large{\textbf{Dynamic Shrinkage Priors for Large Time-varying Parameter Regressions using Scalable Markov Chain Monte Carlo Methods}}\thanks{Corresponding author: Niko Hauzenberger. Department of Economics, University of Salzburg. Address: M\"{o}nchsberg 2A, 5020 Salzburg, Austria. Email: \href{mailto:niko.hauzenberger@plus.ac.at}{niko.hauzenberger@plus.ac.at}. The first two authors gratefully acknowledge financial support by the Austrian Science Fund (FWF): ZK-35 and by funds of the Oesterreichische Nationalbank (Austrian Central Bank, Anniversary Fund, project no. 18127, 18763 and 18765).}}
\author[1, 2]{\MakeUppercase{Niko Hauzenberger}}
\author[1]{\MakeUppercase{Florian Huber}}
\author[2]{\MakeUppercase{Gary Koop}}
\affil[1]{\textit{University of Salzburg}}
\affil[2]{\textit{University of Strathclyde}}

\begin{document}
\maketitle\thispagestyle{empty}\normalsize\small
\begin{center}
\begin{minipage}{0.8\textwidth}
\noindent \textbf{Abstract.} \small Time-varying parameter (TVP) regression models can involve a huge number of coefficients. Careful prior elicitation is required to yield sensible posterior and predictive inferences. In addition, the computational demands of Markov Chain Monte Carlo (MCMC) methods mean their use is limited to  the case where the number of predictors is not too large. In light of these two concerns, this paper proposes a new dynamic shrinkage prior which reflects the empirical regularity that TVPs are typically sparse (i.e., time variation may occur only episodically and only for some of the coefficients). A scalable MCMC algorithm is developed which is capable of handling very high dimensional TVP regressions or TVP Vector Autoregressions. In an exercise using artificial data we demonstrate the accuracy and computational efficiency of our methods. In an application involving the term structure of interest rates in the eurozone, we find our dynamic shrinkage prior to effectively pick out small amounts of parameter change and our methods to forecast well. 
\end{minipage}
\end{center}

\normalsize\renewcommand{\thepage}{\arabic{page}}
\doublespacing
\newpage
\section{Introduction}\label{sec:intro}
\doublespace
The increasing availability of large data sets in economics has led to interest in regressions involving large numbers of explanatory variables. Given the evidence of instability and parameter change in many macroeconomic variables, there is also an interest in time-varying parameter (TVP) regression models and multi-equation extensions such as time-varying parameter Vector Autoregressions (TVP-VARs). This combination of large numbers of explanatory variables with TVPs can lead to regressions with a huge number of parameters. But such regressions are often sparse, in the sense that most of these parameters are zero. In this context, Bayesian methods have proved particularly useful since Bayesian priors can be used to find and impose this sparsity, leading to more accurate inferences and forecasts. A range of priors have been suggested for high-dimensional regression models \citep[see, among many others,][]{ishwaran2005spike, park2008bayesian, griffin2010inference, carvalho2010horseshoe, bhattacharya2015dirichlet}. There is also a growing literature which extends these methods to the TVP case. Examples include \cite{bkk}, \cite{kg2014}, \cite{eisenstat2016stochastic}, \cite{kowal2019dynamic}, \cite{petrova2019quasi}, \cite{kalli2019bayesian}, \cite{bitto2019shrinkage}, \cite{chan2020reducing},   \cite{hauzenberger2022fast} and \cite{fischer2023general}. 
 
Most of these papers assume particular forms of parameter change (e.g., it is common to assume parameters evolve according to random walks) and use computationally-demanding Markov Chain Monte Carlo (MCMC) methods. The former aspect can be problematic (e.g., if parameter change is rare and abrupt, then a model which assumes all parameters evolve gradually according to random walks is inappropriate). The latter aspect means these methods are not scalable (i.e., MCMC-based methods cannot handle models with huge numbers of coefficients).

The contributions of the present paper relate to issues of prior elicitation and computation in TVP regressions. With regards to prior elicitation, we develop novel dynamic shrinkage priors for TVP regressions. These modify recent approaches to dynamic shrinkage priors in papers such as \cite{kowal2019dynamic}. We work with the static representation of the TVP regression model which breaks the coefficients into two groups. One group contains constant coefficients (we call these $\bm \alpha$). The other, which we call $\bm \beta$, are TVPs. In the static representation, the dimension of $\bm \beta$ can be enormous. Our dynamic global-local shrinkage priors are carefully designed to push unimportant elements in $\bm \beta$ to zero in a time-varying fashion. This is done using a global shrinkage parameter that varies over time as well as local shrinkage parameters. The global shrinkage parameter has an interpretation similar to a dynamic factor model with a single factor. This single factor can be used to find periods of time-variation in coefficients and periods when they are constant. Since the assumption of a common volatility factor hampers the use of standard stochastic volatility MCMC algorithms based on a mixture of Gaussians approximation \citep{kim1998stochastic}, we propose a simple approximation that works particularly well in high dimensional settings.

With regards to computation, we develop a scalable MCMC algorithm. This algorithm is suitable for cases where the posterior for $\bm \beta$, conditional on the other parameters in the model, is Gaussian. This occurs for a wide range of global-local shrinkage priors including the dynamic shrinkage priors used in this paper. In this case, the exact MCMC algorithm of \cite{bhattacharya2016fast} is the state of the art.\footnote{{\cite{kastner2020sparse}, \cite{hauzenberger2021flexible} and \cite{korobilis2022new}, for example, use this exact algorithm in the context of large VARs to reduce the computational burden of estimating these models.}} However, even it is too computationally slow to handle the huge number of regressors that appear in the static representation of the TVP regression model. Recently, \cite{johndrow2017bayes} has proposed an approximate algorithm based on this exact algorithm which is computationally much more efficient in sparse models and, thus, is scalable. 

In our paper, it is precisely this scalable MCMC algorithm which forms the basis of the algorithm we use. It involves a thresholding step (described below) which we implement in a different manner than \cite{johndrow2017bayes}. In particular,  as opposed to fixing the threshold to a small number, we set it adaptively. Since this would typically imply a number of thresholds that match the dimension of $\bm \beta$, we  use a method called Signal Adaptive Variable Selection (SAVS), see \cite{bhattacharya2018signal}, to determine the thresholds in a novel way.  SAVS has the advantage of being computationally fast and easy to implement. Recent papers use SAVS for determining variable relevance \citep{hahncarvalho2015dss}, portfolio applications \citep{puelz2020portfolio} or improving macroeconomic forecasts \citep{hko2020}.  We solely use SAVS to identify which variables can be safely set to zero in order to construct an approximate posterior distribution for the TVPs. Thus, the use of SAVS in the context of the algorithm of \cite{johndrow2017bayes} provides two-fold benefits: computational improvements and more flexibility due to its adaptive nature. 

We investigate the use of our methods in artificial and real data. The artificial data exercise demonstrates that our scalable algorithm is a good approximation to exact MCMC and that its computational benefits are substantial. Our application to the eurozone yield curve shows how our methods can effectively pick out small amounts of occasional parameter change in some parameters. Furthermore, allowing for such change in the coefficients improves forecasts. 

The remainder of the paper is organized as follows. The second section defines the TVP regression and TVP-VAR models used in this paper. The third section discusses MCMC methods for the regression coefficients and introduces our computationally-efficient approximate method. Section 4 develops different dynamic shrinkage priors and discusses Bayesian estimation. This section also describes a novel method for drawing the volatilities in the context of a multivariate stochastic volatility process with a common factor. Sections 5 and 6 present our artificial data exercise and our empirical application, respectively. Section 7 summarizes and concludes.   

\clearpage

\section{Static Representation of a TVP Regression}\label{sec:tvp}
\subsection{A TVP regression}
The static representation of a TVP regression model involving a $T$-dimensional dependent variable, $\bm y$, and a $T \times K$-dimensional matrix of predictors, $\bm X$ is:
\begin{equation}
		\bm y = \bm X \bm \alpha + \bm W \bm \beta + \bm L \bm \epsilon, \quad \bm \epsilon \sim \mathcal{N}(\bm 0,\bm I_T),\quad \bm \beta = (\bm \beta'_1, \dots, \bm \beta'_T)',
\label{static}
\end{equation}
where $\bm \alpha$ is a $K$-dimensional vector of time-invariant coefficients, $\bm \beta_t$ is a $K \times 1$ vector of time-varying coefficients and $\bm L = \text{diag}(\sigma_1, \dots, \sigma_T)$ with $\sigma_t$ denoting time-varying error volatilities.
The TVP part of this model arises through the $\bm W \bm \beta$ term. 
 $\bm W$ is a $T \times k (=TK)$ matrix given by:
\begin{equation}
\bm W = \begin{pmatrix} 
  \bm x_1'  & \bm 0_{K \times 1}'   & \dots  & \bm 0_{K \times 1}' \\
  \bm 0_{K \times 1}'    & \bm x_2' & \dots  & \bm 0_{K \times 1}' \\ 
  \vdots    & \vdots   & \ddots & \vdots \\
  \bm 0_{K \times 1}'    & \bm 0_{K \times 1}'   & \dots  & \bm x_T'
\end{pmatrix},
\label{static_w}
\end{equation}
with $\bm x_t$ denoting a $K$-dimensional sub-vector of $\bm X$.  Equation \ref{static} is simply a regression which leads to the terminology \emph{static representation}. But it is a regression with an enormous number of explanatory variables.

Note that  (\ref{static_w}) implies that the TVPs are mean zero and uncorrelated over time.  However, extensions to other forms can be trivially done through a re-definition of $\bm W$. For instance, if we are interested in random walk-type behavior in the TVPs, we can set
\begin{equation}
\bm W = \begin{pmatrix} 
  \bm x_1'  & \bm 0_{K \times 1}'   & \dots  & \bm 0_{K \times 1}' \\
  \bm x_{2}'    & \bm x_2' & \dots  & \bm 0_{K \times 1}' \\ 
  \vdots    & \vdots   & \ddots & \vdots \\
  \bm x_{T}'    & \bm x_{T}'   & \dots  & \bm x_T'
\end{pmatrix}.
\label{rw_w}
\end{equation}
This specification implies that $\bm \beta$ can be interpreted as the changes in the parameters and multiplication with $\bm W$ yields the cumulative sum over $\bm \beta$. In our empirical exercise, we  consider both of these specifications for $\bm W$  and refer to the former as the flexible (FLEX) and the latter as the random walk (RW) specification.  

 The existing literature using Bayesian shrinkage techniques typically uses MCMC methods. Exact MCMC sampling, however, quickly becomes computationally cumbersome since $k$ is extremely large even for moderate values of $T$ and $K$.

Various solutions to this have been proposed in the literature. The standard solution is simply not to work with the static representation, but instead make some parametric assumption about how the TVPs evolve (e.g., assume they follow random walks or a Markov switching process). Unless $K$ is extremely large, exact MCMC methods are feasible. However, with macroeconomic data it is common to find strong evidence of changes in the conditional variance of a series, but much less evidence in favor of change in the conditional mean of a series, \citep[see, e.g.,][]{clark2011}. When $K$ is large, it is plausible to assume that only some of the predictors have time-varying coefficients and, even for these, coefficient change may only rarely happen. Common conventional approaches are not suited for data sets which exhibit such sparsity in the TVPs. If changes in the conditional mean of the parameters happen only rarely then a random walk assumption, which assumes change is continually happening, is not appropriate. If changes in the conditional mean only occur for a small sub-set of the $K$ variables (or occur at different times for different variables), then a Markov switching model which assumes all coefficients change at the same time is not appropriate. These considerations motivate our use of the static representation and the development of a dynamic shrinkage prior suited for the case of TVP sparsity. 

The literature has proposed a few ways of overcoming the computational hurdle that arises if the static representation is used. \cite{korobilis2019high} uses message passing techniques to estimate large TVP regressions and shows that these large models outperform a range of competing models. Similarly, \cite{hkp2020} approximate the TVPs using message passing techniques based on a rotated model representation and sample from the full conditional posterior of $\bm \alpha$ using MCMC methods. Both approaches have the drawback that the quality of the approximation inherent in the use of message passing techniques might be questionable. In another recent paper, \cite{hauzenberger2022fast} propose using the singular value decomposition of $\bm W$ in combination with a conjugate shrinkage prior on $\bm \beta$ to ensure computational efficiency. However, this method has the potential drawback that conjugate priors might be too restrictive for discriminating signals and noise in high dimensional models.  

In this paper, we develop another approach which should work particularly well when $\bm \beta$ is extremely sparse.  This is the scalable MCMC method, based on posterior perturbations, of \cite{johndrow2017bayes}.  
\subsection{Extension to a TVP-VAR}
Before discussing  the scalable MCMC algorithm, we note that methods developed for the TVP regression can also be used for the TVP-VAR if it is written in equation-by-equation form \citep[see, for instance,][]{carriero2019large, hko2020}. In particular, we can use the  following structural representation of the TVP-VAR:
\begin{equation}
	\bm y_t = \bm c_t + \bm A_{0t} \bm y_t + \sum_{p = 1}^{P} \bm A_{pt} \bm y_{t-p} +  \bm \epsilon_t, \quad \bm \epsilon_t \sim \mathcal{N}(0,\bm \Sigma_t),
\label{eq:tvp-var}
\end{equation}
with $\bm y_{t}$ being an $M$-dimensional vector of endogenous variables, $\bm c_t$ denoting an $M$-dimensional vector of intercepts, $\bm A_{pt}$, for $p = 1, \dots, P$, denoting an $M \times M$-dimensional time-varying coefficient matrix that may be stacked in a matrix $\bm A_t = (\bm A_{1t}, \dots, \bm A_{Pt})$.  Furthermore, $ \bm \epsilon_t$ is an $M$-dimensional vector of errors and $\bm \Sigma_t = \text{diag}~(\sigma^2_{1t}, \dots, \sigma^2_{Mt})$ refers to its  diagonal time-varying covariance matrix. Finally, $\bm A_{0t}$ defines contemporaneous relationships between the elements of $\bm y_t$ and is lower-triangular with zeros on the diagonal. 

The $i^{th}~(i=2, \dots, M)$ equation of $\bm y_t$ can be written as a standard TVP regression model: 
\begin{equation*}
y_{it} =  \bm x_{it}'\underbrace{(\bm \alpha_i + \bm \beta_{it})}_{\bm \gamma_{it}} +  \sigma_{it} \epsilon_{it}, \quad \epsilon_{it} \sim \mathcal{N}(0, 1).
\label{eq::eqwse}
\end{equation*}
Here, $\bm x_{it}$ is a $K_i (= MP+i)$-dimensional vector of covariates with $\bm x_{it} = (1, \{y_{jt}\}_{j=1}^{i-1}, \bm y_{t-1}', \dots, \bm y_{t-P}')'$, $\bm \gamma_{it} = (\bm \alpha_i + \bm \beta_{it}) = (c_{it}, \{a_{ij,0t}\}_{j =1}^{i-1}, \bm A_{i\bullet,t})'$ denotes a $K_i$-dimensional vector of time-varying coefficients, with $c_{it}$ referring to the $i^{th}$ element in $\bm c_t$, $a_{ij,0t}$ denoting the $(i,j)^{th}$ element of $\bm A_{0t}$ and $\bm A_{i\bullet,t}$ referring to the $i^{th}$ row of $\bm A_t$. 
For $i=1$, $\bm x_{1t} = (1, \bm y'_{t-1}, \dots, \bm y'_{t-p})'$ and $\bm \gamma_{1t} = (c_{1t},  \bm A_{1\bullet,t})'$. 
Thus, the TVP-VAR can be written as a set of $M$ independent TVP regressions which can be estimated separately using the MCMC methods described in the following section. An additional computational advantage arises in that the $M$ equations can be estimated in parallel using multiple CPUs. 

Depending on the particular choice of $\bm W$, this model nests a variety of commonly used specifications in the literature. For instance, if $\bm W$ implies a random walk behavior of the latent states we arrive at a TVP-VAR closely related to the one proposed in \cite{primiceri2005time}. As we will show below, the main difference is that we have a more flexible state equation by allowing for heteroskedasticity in the shocks to the states through dynamic shrinkage priors. Another model that is closely related to ours is the one proposed in \cite{cogley2010inflation}. This model assumes that the variances of the state innovations evolve according to independent stochastic volatility models. 

\section{Scalable MCMC Algorithm for a Large TVP Model}\label{sec: algorithm}
In this section, we explain the MCMC algorithm of \cite{johndrow2017bayes} and \cite{johndrow2020scalable} and discuss how we adapt it for our TVP regression model. The parameters in the static representation are $\bm \alpha$ and $\bm \beta$. Since $\bm \alpha$ is typically of moderate size and potentially non-sparse, we use conventional (exact) MCMC methods for it. It is $\bm \beta$ which is high-dimensional and potentially sparse, characteristics the algorithm of \cite{johndrow2017bayes} is perfectly suited for. Thus, we use this algorithm for $\bm \beta$. Every model used in the empirical application also includes stochastic volatility. 

In the following section, we develop an MCMC algorithm to produce draws of $\bm L$. Since there is nothing new in our MCMC algorithm for $\bm \alpha$ and our algorithm for drawing $\bm L$ is discussed later, in this section we will proceed conditionally on them and work with the transformed regression involving dependent variable $\tilde{\bm y} = \bm L^{-1} (\bm y - \bm X \bm \alpha)$ and explanatory variables $\tilde{\bm W} = \bm L^{-1} \bm W$. The appendix provides full details of our MCMC algorithm.  In this section, we will also assume that the prior on $\bm \beta$ is (conditional on other parameters) Gaussian with mean zero and a diagonal prior covariance matrix $\bm D_0 = \text{diag}(d_1, \dots, d_{k})$. Many different global-local shrinkage priors have this general form and, in the following section, we will suggest several different choices likely to be well-suited to TVP regressions. 

The exact MCMC algorithm of \cite{bhattacharya2016fast} for drawing $\bm \beta$ proceeds as follows:
\begin{enumerate}
\item Draw a $k$-dimensional vector $\bm v \sim \mathcal{N}(\bm 0_k, \bm D_0)$,
\item Sample a $T$-dimensional vector $\bm q \sim \mathcal{N}(\bm 0_T, \bm I_T)$,
\item Define $\bm w = \tilde{\bm W} \bm v + \bm q$
\item Solve $(\tilde{\bm y} - \bm w) = (\bm I_T + \tilde{\bm W} \bm D_0 \tilde{\bm W}')\bm u$ for $\bm u$,
\item Set $\bm \beta = (\bm D_0 \tilde{\bm W}' \bm u) + \bm v$.
\end{enumerate}
\cite{bhattacharya2016fast} show that this algorithm is fast compared to existing approaches which involve taking the Cholesky factorization of the posterior covariance matrix. However, it can still be slow when $k$ is very large. The computational bottleneck lies in the calculation of $\bm \Gamma = \tilde{\bm W} \bm D_0 \tilde{\bm W}'$ which has computational complexity of order $\mathcal{O}(T^2k)$. In macroeconomic or financial applications involving hundreds of observations, $T^2k = T^3K$ can be enormous.

\cite{johndrow2017bayes} and \cite{johndrow2020scalable} propose an approximation to the algorithm of \cite{bhattacharya2016fast} which, in sparse contexts, will be much faster and, thus, scalable to huge dimensions.
The basic idea of the algorithm is to approximate the high-dimensional matrix $\bm \Gamma$ by dropping irrelevant columns of $\tilde{\bm W}$ so as to speed up computation. To be precise, Steps 4 and 5 of the algorithm are replaced with

\begin{enumerate}
\item[4*] Solve $(\tilde{\bm y} - \bm w) = (\bm I_T +  \hat{\bm \Gamma}) \bm u$ for $\bm u$, with $\hat{\bm \Gamma} = \tilde{\bm W}_S \bm D_{0, S} \tilde{\bm W}_S'$,
\item[5*] Set $\bm \beta = (\bm D_{0, S} \tilde{\bm W}'_S \bm u) + \bm v$.
\end{enumerate}
Here, $\tilde{\bm W}_S$ denotes a $T \times s$-dimensional sub-matrix of $\tilde{\bm W}$ that consists of  columns defined by a set $S$ and $\bm D_{0, S}$ is constructed by taking the diagonal elements of $\bm D_0$ also defined by $S$. Let $S = \{j : {\delta_j} =1\}$ denote an index set with $\delta_j$ being the $j^{th}$ element of a $k$-dimensional selection vector $\bm \delta$ with elements $\delta_j = 1$ with probability $p_j$ and $\delta_j = 0$ with probability $(1-p_j)$. \cite{johndrow2017bayes} approximates $\delta_j$ by setting $\hat{\delta}_j = 0$ if $d_j \in (0, \xi]$ for $\xi$ being a small threshold. Computational complexity is reduced from $\mathcal{O}(T^2k)$  to $\mathcal{O}(T^2 s)$, where $s = \sum_{j=1}^{k} \delta_j$ is the cardinality of the set $S$ or equivalently the number of non-zero parameters in $\bm \beta$. Step 5* yields a draw from the approximate posterior $\hat{p}(\bm \beta | \bullet)$ with the $\bullet$ notation indicating that we condition on the data and the remaining parameters in the model.

The algorithm requires a choice of a threshold for constructing $\bm \delta$. \cite{johndrow2017bayes} suggest simple thresholding rules that seem to work well in their work with artificial data (e.g., recommendations include setting the threshold to $0.01$ when explanatory variables are largely uncorrelated, but $10^{-4}$ when they are more highly correlated). However, choosing the threshold might be problematic for real data applications and can require a significant amount of tuning in practice. Instead we propose to choose the thresholds in a different way using SAVS. 

To explain what SAVS is and how we use it in practice, note first that papers such as \cite{hahncarvalho2015dss} recommend separating out shrinkage (i.e., use of a Bayesian prior to shrink coefficients towards zero) and sparsification (i.e., setting the coefficents on de-selected variables to be precisely zero so as to remove them from the model) into different steps. First, MCMC output from a standard model (e.g., a regression with global-local shrinkage prior) is produced. Secondly, this MCMC output is then sparsified  by choosing a sparse coefficient vector that minimizes the distance between the predictive distribution of the shrunk model and the predictive density of a model based on this sparse coefficient vector plus an additional penalty term for non-zero coefficients. This assumption is critically based on assuming  normally distributed shocks.  The optimal solution, $\tilde{\bm \beta}$, is then a sparse vector which can be used to construct $\bm \delta$. 
 
The advantages of this shrink-then-sparsify approach are discussed in \cite{hahncarvalho2015dss} and, in the context of TVP regressions, in \cite{hko2020}. One important advantage is that estimation error is removed for the sparsified coefficients. When using global shrinkage priors in high dimensional contexts with huge numbers of parameters, small amounts of estimation error can build up and have a deleterious impact on forecasts. By sparsifying, estimation error in the small coefficients is eliminated, thus improving forecasts. This paper differs from the aforementioned papers by using SAVS  to approximate the indicators $\bm \delta$ which is then used in our approximate MCMC algorithm.

The SAVS algorithm, developed in \cite{bhattacharya2018signal}, is a fast method for solving the optimization problem outlined above, making it feasible to sparsify each draw from the posterior of $\bm \beta$.  In the present context, our contention is that a strategy which uses SAVS to shrink-then-sparsify our coefficients can be used to provide a sensible estimate of $\bm \delta$ that does not lead to a deterioration in forecast accuracy. Using SAVS, we first produce a sparsified draws $\tilde{\bm \beta}$.\footnote{Precise details for how SAVS works in TVP regressions, along with additional motivation for the approach, are provided in \cite{hko2020}.} For each draw $\tilde{\bm \beta } = (\tilde{\beta_1}, \dots, \tilde{\beta_{k}})'$, we then set
\begin{equation*}
\hat{\delta}_j = I(\tilde{\beta_j}^* \neq 0). 
\end{equation*}
Each draw of $\hat{\delta}_j$ is used in the construction of $\hat{\bm \Gamma}$ in the MCMC algorithm of \cite{johndrow2017bayes} described above. We will refer to this algorithm as being approximate to distinguish it from the exact algorithm of \cite{bhattacharya2018signal}. 


\section{Bayesian Estimation and Inference}
\subsection{Dynamic global-local shrinkage priors}\label{sec: shrinkagePRIOR}
For the time-invariant coefficients, $\bm \alpha$, we use a horseshoe shrinkage prior \citep{carvalho2010horseshoe}. Since the properties of this prior are familiar and posterior simulation methods for this prior are standard, we do not discuss it further here. See the appendix for additional details. 
 
The important contribution of the present paper lies in the development of a dynamic extension of the horseshoe prior for $\bm \beta $. We modify methods outlined in \cite{kowal2019dynamic}  to design a prior which reflects our beliefs about what kinds of parameter change are commonly found in macroeconomic applications.   In particular, we want to allow for a high degree of sparsity in the TVPs. That is, we want a prior that allows for the possibility that parameter change is rare and may occur for only some coefficients in the regression. There may be periods of instability when parameters change and times of stability when they do not. A dynamic global-local shrinkage prior which has these properties is:
\begin{equation}
p(\bm \beta_t) = \prod_{j=1}^K \mathcal{N}(\beta_{jt}| 0, \tau \lambda_t  \phi^2_{jt}), \quad \phi_{jt} \sim \mathcal{C}^+(0, 1), \label{eq: priordynamic}
\end{equation}
where $\bm \beta_t = (\beta_{1t}, \dots, \beta_{Kt})'$ denotes the coefficients at time $t$, $\tau$ denotes a global shrinkage parameter that pushes all elements in $\bm \beta$ towards zero, $\lambda_t$ is a time-specific shrinkage factor that pushes all elements in $\bm \beta_t$ towards zero and $\phi_{jt}$ is a coefficient and time-specific shrinkage term that follows a half-Cauchy distribution. 

Thus, the prior covariance matrix of $\bm \beta_t$ is given by:
\begin{equation*}
\bm \Omega_t = \tau \lambda_t \times \text{diag}(\phi^2_{1t}, \dots, \phi^2_{Kt}),
\end{equation*}
which implies that $\lambda_t$ acts as a common factor that aims to detect periods characterized by substantive amounts of time variation. 

The main innovation of this paper lies in our treatment of this common factor. Before we discuss the precise specifications for $\lambda_t$, it is worth summarizing the key innovation of this prior. As opposed to the dynamic horseshoe of \cite{kowal2019dynamic}, we only introduce persistence in the common shrinkage factor $\lambda_t$. The key point to note here is that, as opposed to assuming a dynamic law of motion for the coefficient-specific prior scaling parameters, we borrow strength from the cross-sectional dimension and by doing this we substantially reduce the computational burden necessary.

For the global shrinkage parameter we consider  four different laws of motion. The first and second of these involve setting $g_t = \log(\tau \lambda_t)$ and assuming it follows an AR(1) process:
\begin{equation*}
g_t = \mu + \rho (g_{t-1} - \mu) + \nu_t, 
\end{equation*}
with $\mu = \log \tau$. We consider two possible distributions for $\nu_t $. In the first of these it follows a four parameter $Z$-distribution, $\mathcal{Z}(1/2, 1/2, 0, 0)$, leading to a variant of the dynamic horseshoe prior proposed in \cite{kowal2019dynamic} (henceforth labeled \texttt{dHS svol-Z}). The second of these  follows a Gaussian distribution, leading to a standard stochastic volatility model for this prior variance (labeled \texttt{dHS svol-N}).  This model resembles the one stipulated in \cite{cogley2010inflation} but with a single dynamic volatility process. Both of these processes imply a gradual evolution of $g_t$ and thus a  smooth transition from times of rapid parameter change to times of less parameter change.  

The third and fourth specifications allow for more abrupt change between times of stability and times of instability. They assume that $\lambda_t$ is a regime switching process with:
\begin{equation}
\lambda_t =  \kappa_0^2 (1-d_t) + \kappa_1^2 d_t,  \label{eq: mixlambda}
\end{equation}
Here, $d_t$ denotes an indicator that either follows a Markov switching model (labeled \texttt{dHS MS}) or a mixture specification (labeled \texttt{dHS Mix}) and $\kappa_0, \kappa_1$ denote prior variances with the property that $\kappa_1 \gg \kappa_0$. For the Markov switching model, we assume that $d_t$ is driven by a $(2\times2)$-dimensional transition probability matrix $P$ with transition probabilities from state $i$ to  $j$ denoted by $p_{ij}$ (with $p_{ii} \sim \mathcal{B}(a_{i,MS}, b_{i,MS})$, for $i = 0,1$, following a Beta distribution a priori). The mixture model assumes that $p(d_t = 1) = \underline{p}$, with $\underline{p} \sim \mathcal{B}(a_{Mix}, b_{Mix})$.  
In the empirical application we specify $\kappa_1 = 100/K$, $\kappa_0 = 0.01/K$, $a_{Mix} = a_{1,MS} = b_{0,MS} =  3$ and $b_{Mix} = a_{0,MS} = b_{1,MS} = 30$. 

We also include a fifth specification by setting $\lambda_t = 1$ for all $t$. We refer to this setup as the static horseshoe prior (abbreviated as \texttt{sHS}).  For these last three specifications (i.e., the ones that do not assume $\lambda_t$ to evolve according to an AR(1) process), we use a half-Cauchy prior on $\sqrt{\tau} \sim \mathcal{C}^+(0, 1).$

\subsection{Markov Chain Monte Carlo (MCMC) algorithm}
For all of these models, Bayesian estimation and prediction can be done using MCMC methods. 
In this sub-section we mainly focus on how to sample $\lambda_t$ under the assumption that it evolves according to an AR(1) process. For this step we propose a simple and accurate approximation that renders the corresponding hierarchical model linear and conditionally Gaussian. We only briefly discuss the remaining steps since most of them are standard in the literature.

For the time varying regression coefficients, the scalable algorithms (with or without sparsification) of the preceding section, based on \cite{johndrow2017bayes}, can be used. The only modification is that we construct $\bm D_0$ as follows:
\begin{align*}
\bm D_0 = \text{diag}(\bm \Omega_1, \dots, \bm \Omega_T),
\end{align*}
with $\lambda_t$ depending on the specific law of motion adopted. Most of the prior hyperparameters introduced in this section have posterior conditionals of standard forms. These are given in the appendix.  

Sampling $\lambda_t$ for the specifications that assume it to be binary is also  straightforward and can be carried out using standard algorithms.  To sample from the posterior of $\lambda_t$ under the assumption that it evolves according to an AR(1) process, the algorithm proposed in \cite{JPR1995} can be used. However, since this algorithm simulates the $\lambda_t$'s one at a time mixing is often an issue. A second option would be to view the prior  (after squaring each element of $\bm \beta_t$ and taking logs) as the observation equation of a dynamic factor model. This strategy, however, would be computationally challenging for moderate to large values of $K$. As a solution, we propose a new algorithm that is straightforward to implement and, if $K$ is large, has good properties.

Let $\hat{\bm \beta}_t$ be a $K$-dimensional vector of normalized TVPs with typical element $\hat{\beta}_{jt}= {\beta}_{jt}/(\phi_{jt}\tau^{1/2})$. 
Using  (\ref{eq: priordynamic}) and  squaring yields:
\begin{equation}
b_t = (\hat{\bm \beta_t}'\hat{\bm \beta_t}) =  \lambda_t  \nu_t, \label{eq: b(t)}
\end{equation}
with $\nu_t = \bm v'_t \bm v_t$ for $\bm v_t \sim \mathcal{N}(\bm 0_K, \bm I_K)$. Notice that $\nu_t$ follows a $\chi^2$ distribution with $K$ degrees of freedom, denoted by $\chi^2_K$. This implies that sampling algorithms that rely on the Gaussian mixture approximation proposed in \cite{kim1998stochastic} cannot be used. Instead we approximate the  $\chi^2_K$ using a well-known limit theorem that implies, as $K \to \infty$,
\begin{equation*}
\frac{\nu_t - K}{\sqrt{2 K}} \xrightarrow[]{d} \mathcal{N}(0, 1)\quad \Leftrightarrow \quad \nu_t \approx \hat{\nu_t } = \sqrt{2K} q_t + K, \quad q_t \sim \mathcal{N}(0, 1).
\end{equation*}
This approximation works if $K$ is large. In our case, $K$ is often large. For instance, in the largest TVP-VAR model we consider, $K$ is around $100$. Since we estimate the TVP-VAR one equation at a time, values of this order of magnitude hold in each equation and the approximation is likely to be good. But if one were to do full system estimation of the TVP-VAR, there are on the order of $MK$ VAR coefficients at each point in time and the approximation would be even better.   

Substituting the Gaussian approximation into (\ref{eq: b(t)}) and taking logs yields:
\begin{equation}
\log b_t = \log \lambda_t + \log \hat{\nu_t }. \label{eq: approx_2}
\end{equation}
Finally, under the assumption that $(\sqrt{2K}q_t + K)>0$ and by using a Taylor series expansion,\footnote{More precisely, we compute the mean and variance of $\log \hat{\nu_t}$  using a second and first order Taylor series expansion of $\text{E}(\log(K+ \hat{\nu_t}-K))$ and $\text{Var}(\log(K+ \hat{\nu_t}-K))$ around $K$, respectively.} we approximate $\log \hat{v}_t$ with a $\mathcal{N}\left(\log (K)  - 1/K, 2/K\right)$ to render (\ref{eq: approx_2}) conditionally Gaussian. This implies that  any of the standard algorithms proposed in the literature on Gaussian linear state space models can be used. In this paper, we simulate $\log \lambda_t$ using the precision sampler outlined, for example, in \cite{chan2009efficient} and \cite{ MCCAUSLAND2011199}.

\begin{figure}[h!]
\centering
\caption{Approximation error of a single-component Gaussian used to approximate a $\log \chi^2_K$ distribution.}\label{fig: approxLOSS}
\includegraphics[scale=.5]{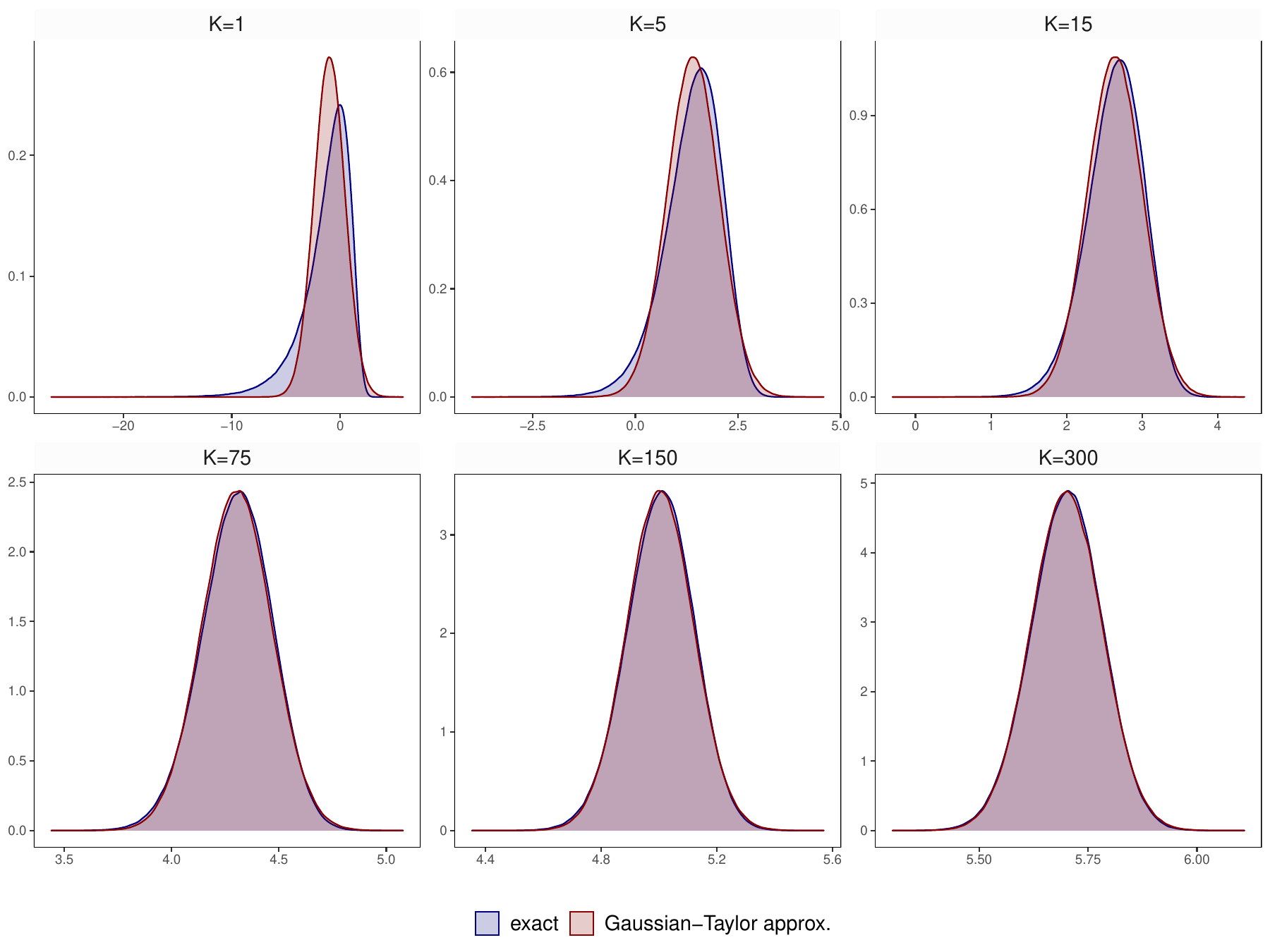}
\caption*{\footnotesize \noindent \textit{Notes}: This figure illustrates the approximation error resulting from approximating the error distribution (which is  $\log \chi^2_K$) with a single-component Gaussian with mean $\log (K) -1/K$ and variance $2/K$. For different values of $K$, the blue shaded areas show the exact error distribution, while the red shaded areas indicate the approximate error distribution.}
\end{figure}

The accuracy of this approximation for different values of $K$ is illustrated in Figure \ref{fig: approxLOSS}. From this figure it is clearly visible that, if $K$ is greater than $5$, our approximation works extremely well. In these cases, there is hardly any difference visible between the $\log \chi^2_K$ and the single-component Gaussian distribution. For $K = 1$ (the most extreme case) and $K=5$, some differences arise which mainly relate to the left tail of the distribution. However, already for $K = 5$ these differences are so small that we do not expect them to have any serious consequences on our estimates of $\lambda_t$, even for small values of $K$.

\section{Illustration Using Artificial Data}
In this section we illustrate the merits of our approach using synthetic data. 
\subsection{How does our algorithm compare to exact MCMC?}
We start by showing that using our approximate (sparsified) algorithm yields estimates that are close to the exact ones in terms of precision. This is achieved by considering five different data generating processes (DGPs). These are all based on Equation (\ref{static}) but make different assumptions about the density and nature of parameter change. Dense DGPs are characterized by having time-variation in a large number of parameters (with sparse DGPs being the opposite of dense). The nature of parameter change can be gradual (e.g., characterized by constant evolution of the parameters) or abrupt.  For each of the five DGPs, we simulate a time series of length $T = 250$ and with $K = 50$. 

The different DGPs assume that the states evolve as follows:
\begin{itemize}
\item \textit{dense gradual:} $\bm \beta_t \sim \mathcal{N}(\bm \beta_{t-1},  \frac{1}{100} \times \bm I_K)$, 
\item \textit{dense mixed:} $\bm \beta_t \sim \mathcal{N}\left(\bm \beta_{t-1}, \left(d_t + \frac{(1-d_tt)}{100}\right) \times \bm I_K\right)$ with $Prob(d_t =1) = 0.1$,
\item \textit{medium-dense gradual:} $\bm \beta_t \sim \mathcal{N}(\bm \beta_{t-1},  \frac{d_t}{100} \times \bm I_K)$ with $Prob(d_t =1) = 0.3$,
\item \textit{sparse abrupt:} $\bm \beta_t \sim \mathcal{N}(\bm \beta_{t-1}, \bm I_K)$ with $Prob(d_t =1) = 0.02$,
\item \textit{no TVPs:} $\bm \beta_t = \bm 0_{K \times 1}$ for all $t$.
\end{itemize}
The remaining parameters are set as follows:   $\bm \beta_0 = \bm 0$, $\bm L = 0.01 \times \bm I_T$, $\bm \alpha \sim \mathcal{N}(\bm 0, \bm I_K)$ and  $\bm X_j \sim \mathcal{N}(\bm 0, \bm I_{T})$ for $j=1,\dots, K$.  Based on these, we use the true path of the parameters $\bm \beta_t$ to obtain a realization of $y_t$. In all simulation experiments and for all models considered we simulate $2,500$ draws from the joint posterior of the parameters and latent states and discard the first $500$ draws as burn-in.

We investigate the accuracy of our scalable approximate MCMC methods relative to the exact MCMC algorithm of \cite{bhattacharya2016fast} (i.e., it is the version of our algorithm which imposes $\delta_j = 1$ for all $j$). Table \ref{tab:sim} shows the ratio of mean absolute errors (MAEs), computed using the posterior mean of $\{\bm \beta_t\}_{t=1}^T$ and the true parameters,
 for the approximate relative to the exact approach for the five priors averaged over the five DGPs. With one exception, MAE ratios are essentially one indicating that the approximate and exact algorithms are producing almost identical results. The one exception is for the DGP which does not have any TVPs. For this case, the approximate algorithm is substantially better than the exact one. This is because our approximate  algorithm uses SAVS which (correctly for this DGP) can set the TVPs to be precisely zero. In this case, draws from the posterior will coincide with draws from the prior that induce heavy shrinkage. Hence, compared to the exact model, the likelihoood does not influence the prior and more shrinkage can be achieved.
 
Thus, Table \ref{tab:sim} shows that, where there is substantial time variation in parameters, the approximation inherent in our scalable MCMC algorithm is an excellent one, yielding results that are virtually identical to the slower exact algorithm. The table also shows the usefulness of SAVS in cases of very sparse DGPs.

\begin{table}[!htbp]
{\tiny
\begin{center}
\caption{Mean absolute errors of the TVPs relative to exact estimation.}\label{tab:sim}
\begin{tabular}{llclllll}
\toprule
\multicolumn{1}{l}{\bfseries }&\multicolumn{1}{c}{\bfseries Specification}&\multicolumn{1}{c}{\bfseries }&\multicolumn{5}{c}{\bfseries MAE ratios: different forms of TVPs}\tabularnewline
\cmidrule{2-2} \cmidrule{4-8}
\multicolumn{1}{l}{}&\multicolumn{1}{c}{}&\multicolumn{1}{c}{}&\multicolumn{1}{c}{dense gradual}&\multicolumn{1}{c}{dense mixed}&\multicolumn{1}{c}{medium-dense gradual}&\multicolumn{1}{c}{sparse abrupt}&\multicolumn{1}{c}{no TVPs}\tabularnewline
\midrule
{\scshape }&&&&&&&\tabularnewline
   ~~&   dHS Mix&   &   1.001&   1.003&   1.001&   1.002&   0.755\tabularnewline
   ~~&   dHS MS&   &   0.998&   0.999&   1.000&   0.999&   0.558\tabularnewline
   ~~&   dHS svol-N&   &   1.000&   1.000&   1.000&   1.000&   0.817\tabularnewline
   ~~&   dHS svol-Z&   &   1.001&   1.001&   1.000&   1.000&   0.696\tabularnewline
   ~~&   sHS&   &   0.999&   1.000&   1.000&   1.001&   0.653\tabularnewline
\bottomrule
\end{tabular}
\begin{minipage}{\textwidth}
\vspace*{10pt}
\footnotesize \textit{Notes:} Numbers are  averages based on $20$ replications from each of the DGPs.  
\end{minipage}
\end{center}}
\end{table}

\subsection{How big are the computational gains of our algorithm?}
Our second artificial data experiment is designed to investigate the computational gains of our algorithm relative to exact MCMC for various choices of $K$, $T$, degrees of sparsity and data configurations. Since we are only interested in computation time we just generate one artificial data set for each of two different ways of specifying $\bm W$. The random numbers refered to below are drawn from the standard Gaussian distribution. 

For $K = 1, ...,400$ and $T \in \{100, 200\}$ we randomly draw a $\bm y$ and an $\bm X$. The $\bm W$ is drawn in two ways which correspond to the flexible and random walk specifications  of equations (\ref{static_w}) and (\ref{rw_w}), respectively. 

In terms of sparsity, we consider four scenarios based on how we choose $\tilde{\bm W}_S$:
\begin{itemize}
\item $100\%$ dense: $\tilde{\bm W}_S = \bm W$. This is the exact algorithm.
\item $50\%$ dense:  $\tilde{\bm W}_S $ contains $50\%$ of the columns of $\bm W$ (i.e., $s = 0.5k$).
\item $10\%$ dense:  $\tilde{\bm W}_S $ contains $10\%$ of the columns of $\bm W$ (i.e., $s = 0.1k$).
\item $1\%$ dense: $\tilde{\bm W}_S $ contains $1\%$ of the columns of $\bm W$ (i.e., $s = 0.01k$).
\end{itemize}

\begin{figure}[!h]
\centering
\caption{Time necessary to obtain a draw for $K$ time-varying coefficients.}\label{fig:comput}
\includegraphics[scale=0.52]{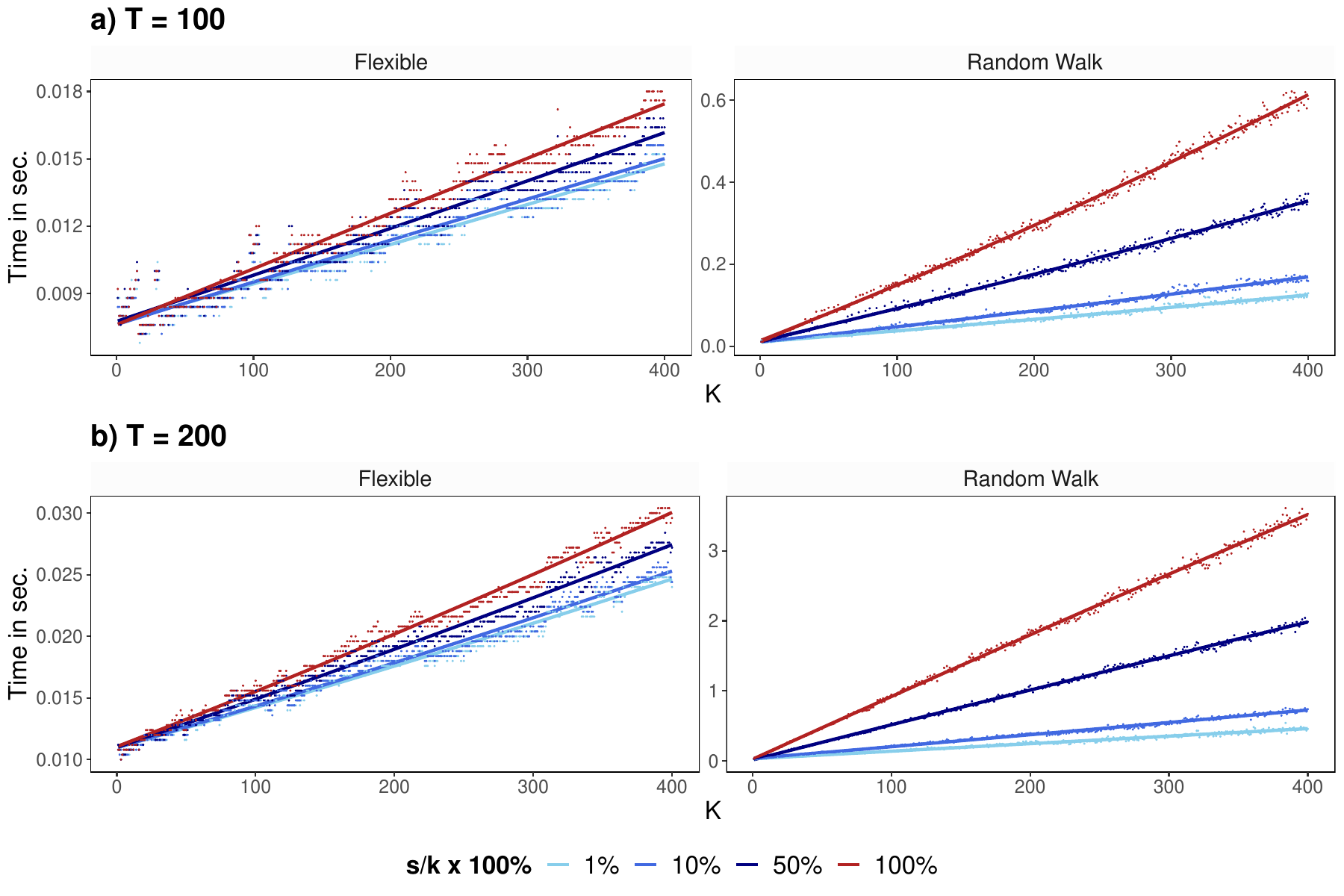}
\caption*{\footnotesize \noindent \textit{Notes}: The figure shows the estimation time in seconds required to obtain a draw for $K$ time-varying coefficients for different degrees of overall sparsity (i.e., $1\%, 10\%, 50\%$, and $100\%$ dense). The dots refer to the empirical run times for which we fit a nonlinear trend (indicated by the solid lines). The red colored dots and red solid lines indicate run times of the exact algorithm ($100\%$ dense, with $s = k$).}
\end{figure}

Figure \ref{fig:comput} depicts the computational advantages of our approximate MCMC algorithm relative to the exact algorithm of \cite{bhattacharya2016fast}. It shows the time necessary to obtain a draw of $\bm \beta$. It can be seen that when the TVPs are highly correlated over time as with the random walk specification, then our scalable algorithm has substantial computational advantages relative to the exact algorithm particularly for large $K$ and in sparse data sets. When the TVPs are uncorrelated the computational advantages of our approach relative to the exact algorithm are smaller, but still appreciable.\footnote{The relatively good performance of the exact algorithm in this case is partly due to the fact that we are coding using sparse algorithms. In the flexible specification for $\bm W$, the underlying matrices are block-diagonal and thus exact sampling is already quite fast.}
 
\section{Empirical Application using Eurozone Yield Data}
\subsection{Data overview and specification issues}
We illustrate our methods using a monthly data set of $30$ government bond yields in the euro area (EA).  As opposed to forecasting standard US macroeconomic time series such as output, inflation and unemployment rates, forecasting EA government bond yields is challenging due to, at least, three reasons. The first is that the researcher has to decide on the segment of yield curve she is interested in or use techniques that allow for analyzing the full term structure of government bond yields. Following the latter approach leads to overfitting issues whereas the former approach might suffer from omitted variable bias. The second challenge is that these time series are often subject to outliers as well as sharp shifts in the conditional variance.  The final reason is that the time series we consider are rather short and in such circumstances TVP-VARs risk overfitting if the estimates of the TVPs are not regularized sufficiently. We expect that the techniques proposed in this paper are capable of handling both issues  well.

We use monthly yield curve data obtained from Eurostat.  This dataset includes the yield to maturity of a (hypothetical) zero coupon bond on AAA-rated government bonds of eurozone countries for $30$ different maturities. These maturities range from one-year to $30$-years and span the period from $2005$:$01$ to $2019$:$12$. 

If we wish to model all $30$ yields jointly we have to estimate a TVP-VAR with $M=30$ equations, a challenging statistical and computational task which we will take on in the next sub-section. Since the parameter space of such a model is vast and difficult to interpret, in this sub-section where we present some in-sample results, we will use a small-scale example. This model is based on the Nelson-Siegel three factor model \citep[see, e.g., ][]{nelson1987parsimonious, diebold2006macroeconomy} and assumes that the yield on a security with maturity $\mathfrak{t}$, labeled $r_t(\mathfrak{t})$, features a factor structure:
\begin{equation}
r_{t}(\mathfrak t) = L_{t} + S_{t}  \left( \frac{1-e^{-\zeta \mathfrak t}}{\zeta \mathfrak t}\right)+C_{t} \left( \frac{1-e^{-\zeta \mathfrak{t}}}{\zeta \mathfrak{t}} - e^{-\zeta \mathfrak{t}} \right) + \eta_t(\mathfrak t), \quad \eta_t(\mathfrak t) \sim \mathcal{N}\left(0, \sigma^2_\eta(\mathfrak t)\right) . \label{eq: obs_NS}
\end{equation}
Here,  $L_{t}, S_{t}$ and $C_{t}$ refer to the level, slope and curvature factor, respectively, while $\eta_t(\mathfrak t)$ denotes maturity-specific measurement errors which are independent across maturities and feature variance $\sigma^2_\eta(\mathfrak t)$. $\zeta$ denotes a parameter that controls the shape of the factor loadings. Following \cite{diebold2006macroeconomy}, we set $\zeta = 0.7308~(12 \times 0.0609)$. Since the loading of the level factor is one for all maturities and does not feature a discount factor, it defines the behavior at the long end of the yield curve. Moreover, the slope factor mainly shapes the short end of the yield curve and the curvature factor defines the middle part of the curve. 
The latent yield curve factors are obtained by running OLS on a $t$-by-$t$ basis. These estimates are then consequently used as our endogenous variables by  setting $\bm y_{t} = (L_{t}, S_{t}, C_{t})'$ and estimating the TVP-VAR defined in (\ref{eq:tvp-var}). We use the flexible specification for $\bm W$ in (\ref{static_w}) and the approximate algorithm to estimate the model. In addition, we set the lag length to two. After obtaining forecasts for $\bm y_t$, we use (\ref{eq: obs_NS}) to map the factors back to the observed yields. It is worth noting that (\ref{eq: obs_NS}) constitutes an observation equation which links the observed yields to the latent Nelson-Siegel factors. To compute predictive densities, we also take the corresponding measurement errors into account by estimating the measurement error variance independently for each observed series.

\subsection{In-sample results}
\begin{figure}
\centering
\caption{Heatmaps of posterior inclusion probability (PIPs) for time-variation in structural TVP-VAR coefficients with a gradually changing common shrinkage factor.}\label{fig:coeffs1}
\begin{minipage}{\textwidth}
\centering
a) \textit{\texttt{dHS svol-Z}}
\vspace{5pt}
\end{minipage}
\begin{minipage}{\textwidth}
\centering
\includegraphics[scale=0.64]{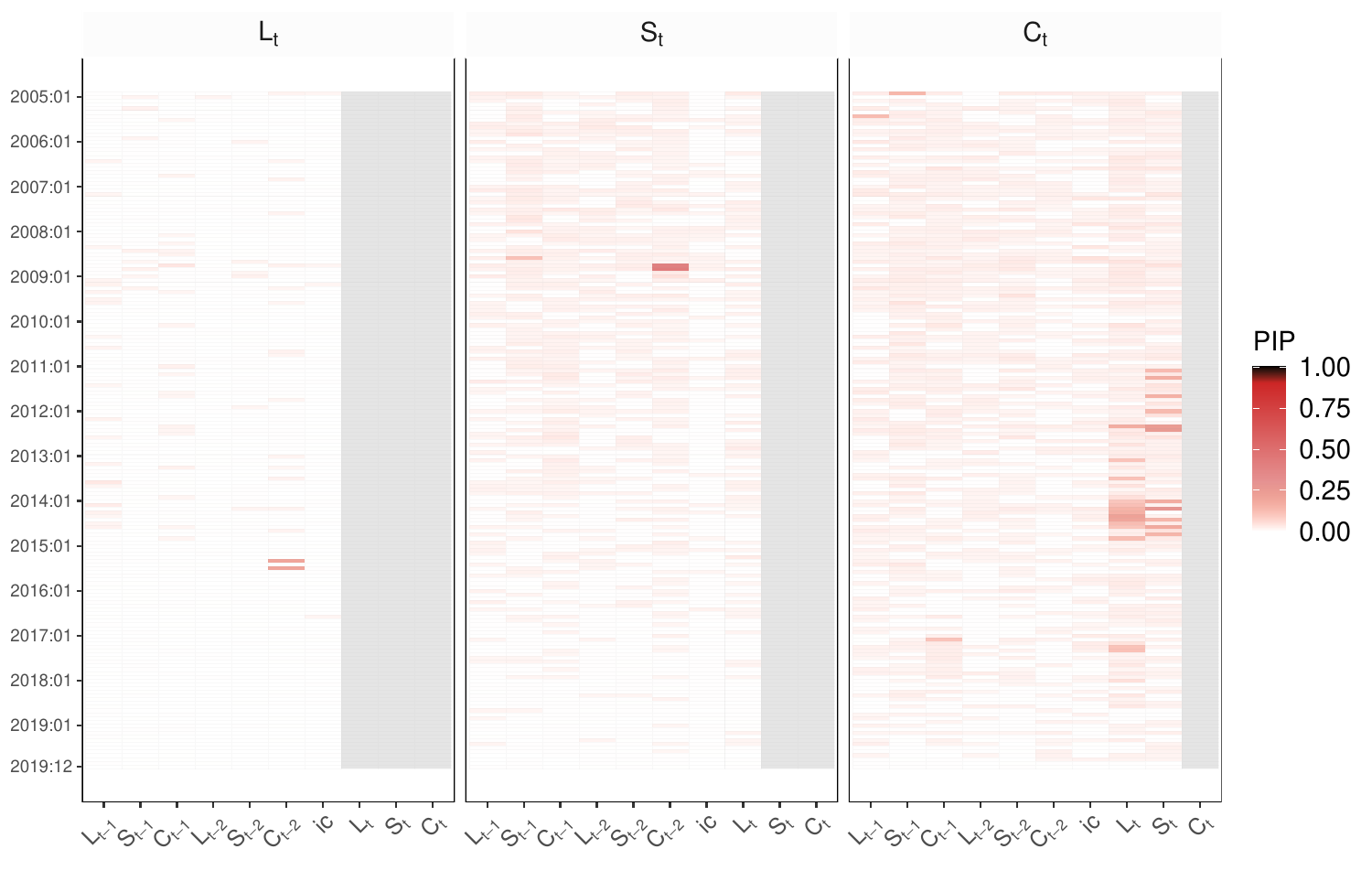}
\end{minipage}
\begin{minipage}{\textwidth}
\vspace{10pt}
\centering
b) \textit{\texttt{dHS svol-N}}
\vspace{5pt}
\end{minipage}
\begin{minipage}{\textwidth}
\centering
\includegraphics[scale=0.64]{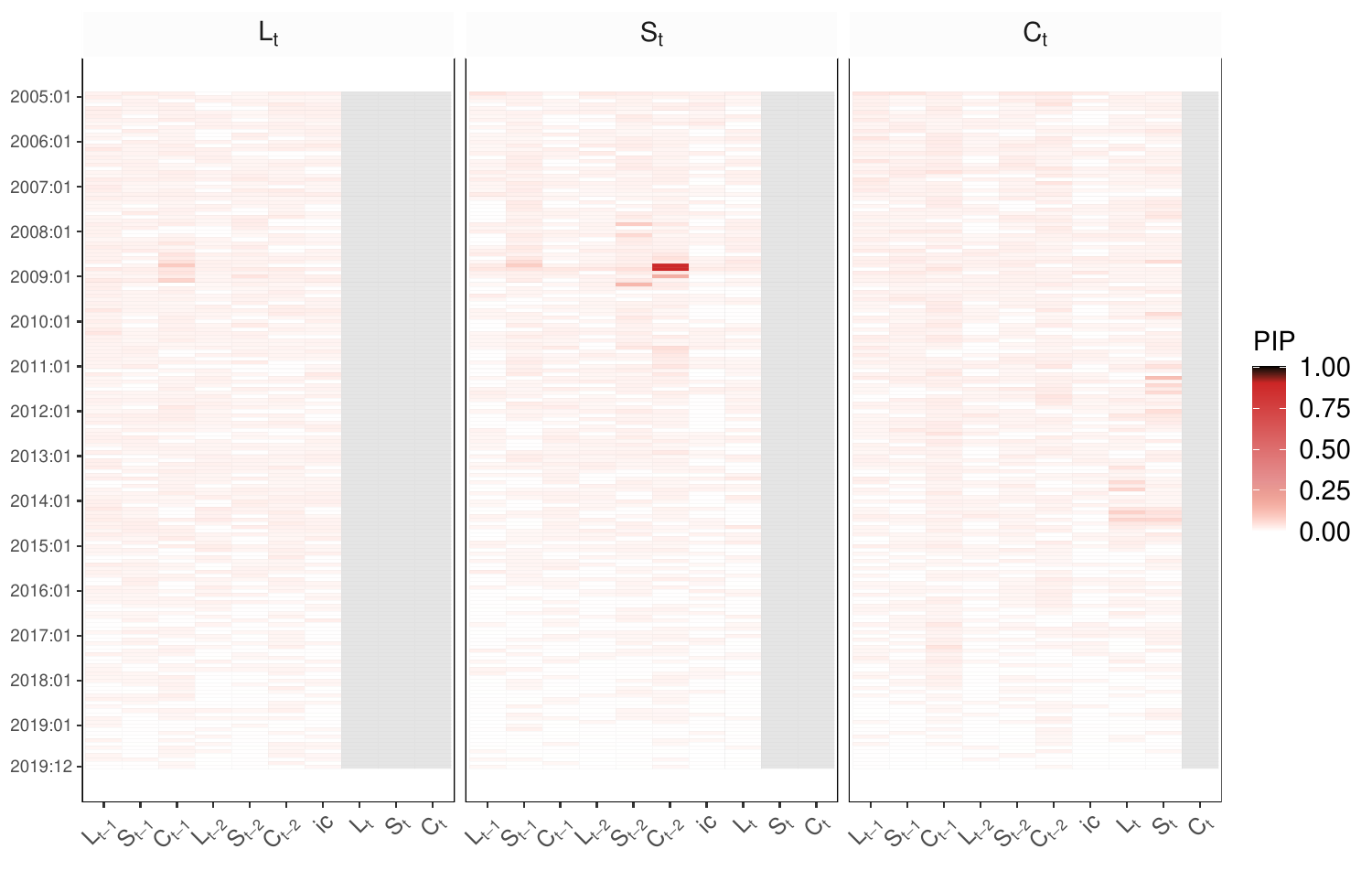}
\end{minipage}\hfill
\caption*{\footnotesize \noindent \textit{Notes}: Grey shaded areas indicate coefficients which do not appear in the model due to the lower triangularity of $\bm A_{0t}$.}
\end{figure}

\begin{figure}
\centering
\caption{Heatmaps of posterior inclusion probability (PIPs) for time-variation in structural TVP-VAR coefficients with a regime-switching common shrinkage factor.}\label{fig:coeffs2}
\begin{minipage}{\textwidth}
\centering
a) \textit{\texttt{dHS Mix}}
\vspace{5pt}
\end{minipage}
\begin{minipage}{\textwidth}
\centering
\includegraphics[scale=0.64]{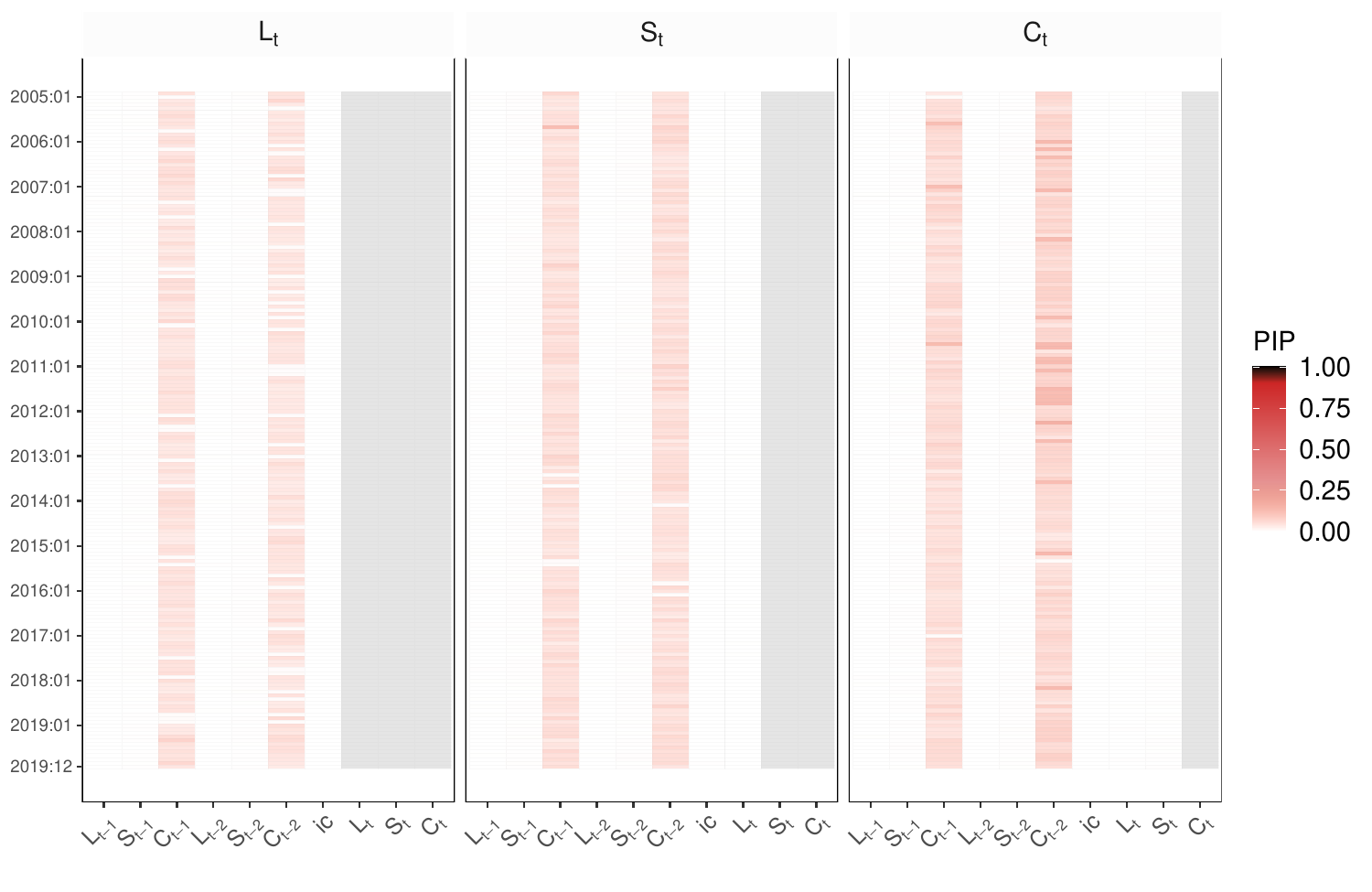}
\end{minipage}
\begin{minipage}{\textwidth}
\vspace{10pt}
\centering
b) \textit{\texttt{dHS MS}}
\vspace{5pt}
\end{minipage}\hfill
\begin{minipage}{\textwidth}
\centering
\includegraphics[scale=0.64]{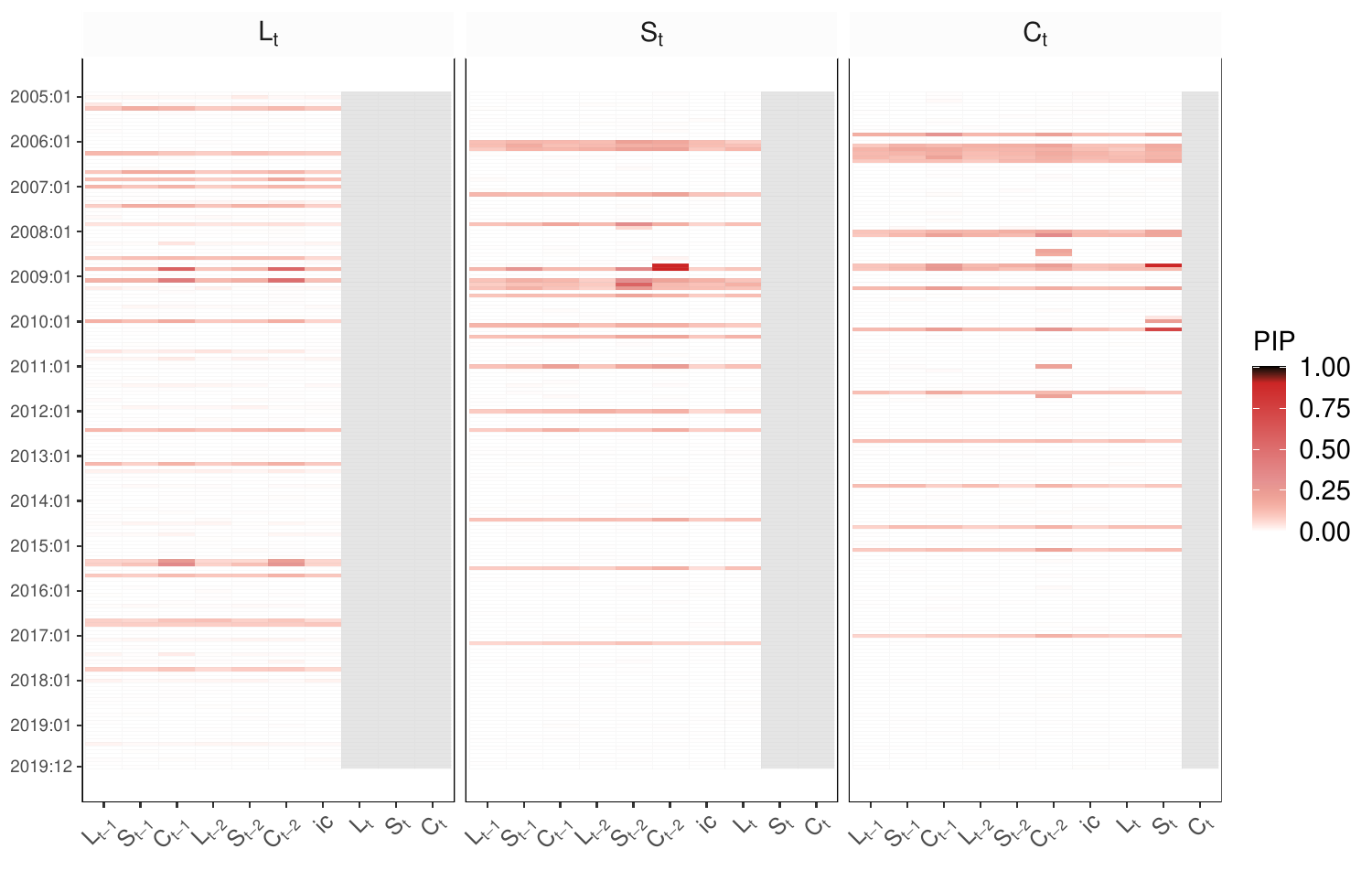}
\end{minipage}\hfill
\caption*{\footnotesize \noindent \textit{Notes}: Grey shaded areas indicate coefficients which do not appear in the model due to the lower triangularity of $\bm A_{0t}$.}
\end{figure}

To provide some information on the amount of time variation, Figures \ref{fig:coeffs1} and \ref{fig:coeffs2} depicts heatmaps of the posterior inclusion probability (PIPs) for a Nelson-Siegel model with panels a) to d) referring to the four different dynamic priors for $\lambda_t$.  These PIPs are the posterior means of the elements of $\bm \delta$.

The main impression provided by Figure \ref{fig:coeffs1} and \ref{fig:coeffs2} is that there is little evidence of strong time-variation in the parameters when using this data set. However, there does seem to be some in the sense that there are many variables and time periods where the PIPs are appreciably above zero. That is, even though the figures contain a lot of white (PIPs essentially zero) and just a handful of deep reds (PIPs above one half), there is a great deal of pink of various shades (e.g., PIPs 20\%-30\%). This is consistent with time-variation being small, episodic and only occurring in some coefficients. 

Results for our four different dynamic horseshoe priors are slightly different indicating the dynamic prior choice can have an impact on results. A clear pattern emerges only for the dynamic horseshoe prior with a mixture specification. It is finding that small amounts of time-variation occur only for the coefficients on the curvature factor. If the mixture part of the prior is replaced by a Markov switching specification, we tend to find short-lived periods where a small amount of time-variation occurs for all of the coefficients in an equation. But, interestingly, \texttt{dHS MS} finds that different equations have time-variation occuring at different periods of time. Evidence for TVPs is the least when we use stochastic volatility specifications in the dynamic horseshoe priors. For these priors, tiny amounts of time variation (i.e., tiny PIPs) are spread much more widely throughout the sample and across variables.

\subsection{Forecast exercise}
The dataset covers the entire yield curve and includes yields from one-year to thirty-year bonds in one-year steps. We choose $\{1$y, $3$y, $5$y, $7$y, $10$y, $15$y, $30$y$\}$ maturities as our target variables that we wish to forecast and consider one-month and one-quarter ahead as forecast horizons. We use a range of competing models that differ in terms of how they model time-variation in coefficients and the number of endogenous variables they have. All models feature stochastic volatility in the measurement errors and have two lags. We also offer comparison between the two MCMC algorithms: exact and approximate. 

In terms of VAR dimension, we have large TVP-VARs and VARs with all 30 maturities ($M=30$) as well as the three factor Nelson-Siegel model described in the previous sub-section ($M=3$). 

In terms of time variation specified through the likelihood function (i.e., through the definition of $\bm W$), we consider the flexible (FLEX) and random walk (RW) specifications defined in (\ref{static_w}) and (\ref{rw_w}). In terms of time variation specified through the prior, we consider the five global-local shrinkage priors (four dynamic and one static)  given in Sub-section \ref{sec: shrinkagePRIOR}. In addition, we consider as a competitor the conventional TVP-VAR setup of \cite{primiceri2005time}. We estimate the TVP-VAR only for the Nelson-Siegel model since the original prior overfits in higher dimensions.\footnote{The priors of the conventional \cite{primiceri2005time} TVP-VAR are informed by OLS estimates using an initial training sample (in our case the initial first $18$ observations). Such an empirical Bayesian calibration strategy is only sensible for models that feature a small number of endogenous variables.}

We also have VAR models where coefficients are constant over time. For these we do two versions, one with a Minnesota prior (MIN) and the other a horseshoe prior (HS).  These models are estimated by setting $\bm \beta = \bm 0$ and then using the sampling steps for $\bm \alpha$ detailed in the appendix. For the Minnesota prior, we use a non-conjugate version that allows for asymmetric shrinkage patterns and integrate out the corresponding hyperparameters within MCMC.


To evaluate one-month and one-quarter ahead forecasts, we use a recursive prediction design and split the sample into an initial estimation period that ranges from $2005$:$01$ to $2008$:$12$ and a forecast evaluation period from $2009$:$01$ to $2019$:$12$.  We use Root Mean Squared Forecast Errors (RMSEs) as the measure of performance for our point forecasts and Continuous Ranked Probability Scores \citep[CRPSs,][]{gneiting2007strictly} as the measure of performance of our density forecasts. Both are presented in ratio form relative to the benchmark model which is the large VAR with Minnesota prior. Values less than one indicate an approach is beating the benchmark. 

We present our forecasting results in two tables. Table \ref{tab:non-sps1} shows the one-month ahead forecast performance of the different models while Table \ref{tab:non-sps3} in the appendix shows the one-quarter ahead forecasting results. Our focus on one-step ahead forecasts is predicated by the fact that the density forecast measures based on proper scoring rules (such as CRPSs) can be viewed as a training sample marginal likelihood and thus enables model comparison \citep[see][]{gneiting2007strictly}.

Overall, the evidence in Table \ref{tab:non-sps1} (and Table \ref{tab:non-sps3}) is mixed, with no single approach being dominant. In principle, one robust pattern is that models with TVPs tend to produce more accurate forecasts than the large VAR with stochastic volatility benchmark. These gains range from being rather small (particularly at the short-end of the yield curve) to appreciable (when the focus is on the long-end of the yield curve). This is consistent with recent findings in \cite{fischer2023general} who document that flexible models work well for this particular dataset when longer maturities are considered.

If we compare results for the large TVP-VARs to results for the smaller TVP-VARs based on the Nelson-Siegel factors reveals that both specifications produce forecasts of similar quality. When forecasting one-month ahead and focusing on the CRPS as a measure of forecast performance, the best average forecast performance is produced by one of the large TVP-VARs. But when we focus on point forecasting performance, one of the NS models emerges as the best forecasting model. This finding indicates that using more information in an unrestricted manner seems to exert benign effect on higher order moments of the predictive density whereas for the first moment the effect is negligible (or even negative). Interestingly, this finding only holds for one-month ahead predictive densities. When we focus on one-quarter ahead forecasts (see Table \ref{tab:non-sps3} in the appendix), this result is reversed with CRPSs indicating one of the NS models is forecasting best and RMSEs indicating one of the large TVP-VARs is forecasting best. 

The comparison of the different choices for $\bm W$ also yields a mixed pattern of results. At the short end of the yield curve the RW specification tends to forecast better, but at the longer end the FLEX specification does better. It is interesting to note, however, that the good performance for RW occurs with a large TVP-VAR whereas for the FLEX specification it occurs for a  Nelson-Siegel version of the model. 

In terms of which of our dynamic horseshoe priors forecasts best, it does seem to be the priors which assume $\lambda_t$ to exhibit rapid change between values forecast better than the gradual change of the stochastic volatility specifications. That is, the Markov switching or mixture versions of the prior, \texttt{dHS MS} and \texttt{dHS Mix}, tend to forecast better than \texttt{dHS svol-Z} or \texttt{dHS svol-N}. Although there are several exceptions to this pattern. At this point it is also worth highlighting that the original \cite{primiceri2005time} model is outperformed by our shrinkage specifications in all segments of the yield curve. This suggests that using proper shrinkage priors on the state innovation variances and allowing for dynamic shrinkage pays off.

Thus, overall (and with several exceptions) we have a story where, in this data set, time variation in the regression coefficients is present and there are gains to be made from capturing them. This can be seen by noting that the constant parameter VARs with stochastic volatility are never the best performing specifications across the different maturities and also for both time horizons we consider. As we have shown in the previous sub-section, this time variation is episodic (rather than gradually evolving) and only occurs occasionally and for some of the coefficients. However, ignoring this time variation and using constant parameter models leads to a  deterioration in forecasts in almost all situations. 

In terms of computation, our scalable algorithm does seem to work well. If we compare results from the exact MCMC algorithm to our approximate (non-sparsified) algorithm, it can be seen that using the computationally-faster approximation is not leading to a deterioration in forecast performance. In fact, there are some cases where the approximate forecasts are better than their exact counterparts. 
\clearpage
{\tiny
\begin{longtable}{llcllllllll}
\caption{One-month ahead forecast performance for EA central government bond yields at different maturities using non-sparsified models. 
\label{tab:non-sps1}}\tabularnewline
\toprule
\multicolumn{1}{l}{\bfseries }&\multicolumn{1}{c}{\bfseries Specification}&\multicolumn{1}{c}{\bfseries }&\multicolumn{8}{c}{\bfseries One-month-ahead}\tabularnewline
\cmidrule{4-11}
\multicolumn{1}{l}{}&\multicolumn{1}{c}{}&\multicolumn{1}{c}{}&\multicolumn{1}{c}{Avg.}&\multicolumn{1}{c}{1y}&\multicolumn{1}{c}{3y}&\multicolumn{1}{c}{5y}&\multicolumn{1}{c}{7y}&\multicolumn{1}{c}{10y}&\multicolumn{1}{c}{15y}&\multicolumn{1}{c}{30y}\tabularnewline
\midrule
\endfirsthead
\multicolumn{11}{r}{\footnotesize{\textit{Table \ref{tab:non-sps1} continued}}}\tabularnewline
\toprule
\multicolumn{1}{l}{\bfseries }&\multicolumn{1}{c}{\bfseries Specification}&\multicolumn{1}{c}{\bfseries }&\multicolumn{8}{c}{\bfseries One-month-ahead}\tabularnewline
\cmidrule{4-11}
\multicolumn{1}{l}{}&\multicolumn{1}{c}{}&\multicolumn{1}{c}{}&\multicolumn{1}{c}{Avg.}&\multicolumn{1}{c}{1y}&\multicolumn{1}{c}{3y}&\multicolumn{1}{c}{5y}&\multicolumn{1}{c}{7y}&\multicolumn{1}{c}{10y}&\multicolumn{1}{c}{15y}&\multicolumn{1}{c}{30y}\tabularnewline
\midrule
\endhead
\multicolumn{11}{c}{\textbf{VAR with constant coefficients}}\tabularnewline
\shadeBench   ~~&   Large with MIN&   &   0.99&   0.70&   0.84&   0.90&   0.96&   1.04&   1.16&   1.23\tabularnewline
\shadeBench   ~~&   &   &   (0.51)&   (0.34)&   (0.43)&   (0.50)&   (0.54)&   (0.57)&   (0.60)&   (0.60)\tabularnewline
   ~~&   Large VAR with HS prior&   &   0.93&   0.98&   0.96&   0.97&   0.98&   0.96&   0.90&   0.82\tabularnewline
   ~~&   &   &   (0.96)&   (0.98)&   (0.97)&   (0.98)&   (0.98)&   (0.97)&   (0.95)&   (0.90)\tabularnewline
   ~~&   Nelson-Siegel VAR with HS prior&   &   0.91&   1.02&   0.98&   0.97&   0.97&   0.94&   0.88&   0.78\tabularnewline
   ~~&   &   &   (0.96)&   (1.01)&   (1.01)&   (0.99)&   (0.98)&   (0.96)&   (0.94)&   (0.87)\tabularnewline
   ~~&   Nelson-Siegel with MIN prior&   &   0.92&   1.01&   0.99&   0.98&   0.97&   0.94&   0.88&   0.78\tabularnewline
   ~~&   &   &   (0.97)&   (1.03)&   (1.03)&   (1.01)&   (0.99)&   (0.96)&   (0.94)&   (0.87)\tabularnewline
\midrule
\multicolumn{11}{c}{\textbf{Large TVP-VAR with the random walk specification for $\bm W$}}\tabularnewline
   ~~&   dHS Mix&   &   0.98&   1.00&   0.97&   1.01&   1.04&   1.03&   0.96&   0.92\tabularnewline
   ~~&   &   &   (1.00)&   (0.99)&   (0.98)&   (0.99)&   (1.01)&   (1.01)&   (1.00)&   (0.98)\tabularnewline
\shadeRow   ~~&   dHS Mix (approx.)&   &   0.93&   0.99&   0.98&   0.99&   0.99&   0.95&   0.90&   0.83\tabularnewline
\shadeRow   ~~&   &   &   (0.97)&   (0.97)&   (0.98)&   (0.98)&   (0.98)&   (0.97)&   (0.97)&   (0.94)\tabularnewline
   ~~&   dHS MS&   &   0.97&   0.97&   0.97&   1.00&   1.02&   1.01&   0.96&   0.87\tabularnewline
   ~~&   &   &   (0.99)&   (0.98)&   (0.98)&   (1.00)&   (1.01)&   (1.01)&   (1.00)&   (0.95)\tabularnewline
\shadeRow   ~~&   dHS MS (approx.)&   &   0.91&   0.97&   0.95&   \textbf{0.96}&   \textbf{0.96}&   0.94&   0.88&   0.82\tabularnewline
\shadeRow   ~~&   &   &   (\textbf{0.95})&   (0.98)&   (\textbf{0.96})&   (\textbf{0.96})&   (\textbf{0.97})&   (0.96)&   (0.95)&   (0.92)\tabularnewline
   ~~&   dHS svol-N&   &   0.92&   0.99&   0.98&   0.98&   0.98&   0.94&   0.88&   0.81\tabularnewline
   ~~&   &   &   (0.96)&   (0.98)&   (0.99)&   (0.99)&   (0.98)&   (0.97)&   (0.95)&   (0.89)\tabularnewline
\shadeRow   ~~&   dHS svol-N (approx.)&   &   0.93&   0.98&   0.97&   0.99&   0.99&   0.96&   0.90&   0.82\tabularnewline
\shadeRow   ~~&   &   &   (1.01)&   (0.98)&   (0.98)&   (0.99)&   (0.99)&   (1.03)&   (1.06)&   (1.02)\tabularnewline
   ~~&   dHS svol-Z&   &   0.94&   0.98&   0.96&   0.98&   0.99&   0.97&   0.92&   0.83\tabularnewline
   ~~&   &   &   (0.99)&   (0.97)&   (0.97)&   (0.99)&   (0.99)&   (0.99)&   (0.97)&   (1.02)\tabularnewline
\shadeRow   ~~&   dHS svol-Z (approx.)&   &   0.93&   0.98&   0.97&   0.99&   0.99&   0.97&   0.90&   0.83\tabularnewline
\shadeRow   ~~&   &   &   (0.97)&   (0.98)&   (0.98)&   (0.99)&   (0.99)&   (0.98)&   (0.96)&   (0.92)\tabularnewline
   ~~&   sHS&   &   0.96&   1.00&   0.96&   0.99&   1.02&   1.01&   0.95&   0.88\tabularnewline
   ~~&   &   &   (1.04)&   (0.99)&   (0.97)&   (0.99)&   (1.02)&   (1.02)&   (1.13)&   (1.09)\tabularnewline
\shadeRow   ~~&   sHS (approx.)&   &   0.93&   0.99&   \textbf{0.94}&   0.97&   0.98&   0.96&   0.91&   0.82\tabularnewline
\shadeRow   ~~&   &   &   (0.97)&   (0.98)&   (0.97)&   (0.98)&   (0.98)&   (0.98)&   (0.97)&   (0.92)\tabularnewline
\midrule
\multicolumn{11}{c}{\textbf{Large TVP-VAR with the flexible specification for $\bm W$}}\tabularnewline
   ~~&   dHS Mix&   &   1.05&   0.99&   0.96&   1.01&   1.07&   1.09&   1.07&   1.05\tabularnewline
   ~~&   &   &   (1.04)&   (0.98)&   (0.98)&   (1.02)&   (1.05)&   (1.07)&   (1.08)&   (1.06)\tabularnewline
\shadeRow   ~~&   dHS Mix (approx.)&   &   0.91&   0.98&   0.96&   0.97&   0.97&   0.94&   0.89&   0.78\tabularnewline
\shadeRow   ~~&   &   &   (1.01)&   (0.98)&   (0.97)&   (0.98)&   (0.98)&   (0.98)&   (0.99)&   (1.18)\tabularnewline
   ~~&   dHS MS&   &   1.13&   1.00&   1.03&   1.12&   1.17&   1.17&   1.17&   1.14\tabularnewline
   ~~&   &   &   (1.18)&   (1.04)&   (1.10)&   (1.16)&   (1.20)&   (1.21)&   (1.22)&   (1.26)\tabularnewline
\shadeRow   ~~&   dHS MS (approx.)&   &   0.93&   0.98&   0.97&   0.99&   0.99&   0.95&   0.89&   0.82\tabularnewline
\shadeRow   ~~&   &   &   (1.01)&   (0.98)&   (0.98)&   (0.99)&   (0.99)&   (0.97)&   (0.96)&   (1.15)\tabularnewline
   ~~&   dHS svol-N&   &   0.92&   0.98&   0.95&   0.96&   0.97&   0.95&   0.89&   0.85\tabularnewline
   ~~&   &   &   (0.96)&   (0.98)&   (0.97)&   (0.97)&   (0.98)&   (0.97)&   (0.95)&   (0.91)\tabularnewline
\shadeRow   ~~&   dHS svol-N (approx.)&   &   0.92&   0.98&   0.95&   0.97&   0.97&   0.95&   0.91&   0.83\tabularnewline
\shadeRow   ~~&   &   &   (1.00)&   (0.98)&   (0.97)&   (0.97)&   (0.97)&   (0.97)&   (0.96)&   (1.15)\tabularnewline
   ~~&   dHS svol-Z&   &   0.95&   0.98&   0.96&   0.98&   1.00&   0.98&   0.92&   0.88\tabularnewline
   ~~&   &   &   (0.98)&   (0.98)&   (0.98)&   (0.99)&   (1.00)&   (1.00)&   (0.99)&   (0.95)\tabularnewline
\shadeRow   ~~&   dHS svol-Z (approx.)&   &   0.93&   0.97&   0.96&   0.98&   0.98&   0.96&   0.92&   0.83\tabularnewline
\shadeRow   ~~&   &   &   (0.96)&   (0.98)&   (0.97)&   (0.98)&   (0.98)&   (0.97)&   (0.96)&   (0.91)\tabularnewline
   ~~&   sHS&   &   0.92&   0.98&   0.96&   0.97&   0.97&   0.94&   0.89&   0.82\tabularnewline
   ~~&   &   &   (0.95)&   (0.98)&   (0.97)&   (0.98)&   (0.97)&   (0.96)&   (0.95)&   (0.89)\tabularnewline
\shadeRow   ~~&   sHS (approx.)&   &   0.93&   0.96&   0.95&   0.97&   0.98&   0.96&   0.91&   0.83\tabularnewline
\shadeRow   ~~&   &   &   (0.96)&   (\textbf{0.96})&   (0.97)&   (0.98)&   (0.98)&   (0.97)&   (0.95)&   (0.92)\tabularnewline
\midrule
\multicolumn{11}{r}{}\tabularnewline
\multicolumn{11}{r}{}\tabularnewline
\multicolumn{11}{r}{}\tabularnewline
\multicolumn{11}{c}{\textbf{Nelson-Siegel TVP-VAR with the random walk specification for $\bm W$}}\tabularnewline
   ~~&   dHS Mix&   &   1.11&   1.14&   1.21&   1.19&   1.18&   1.15&   1.08&   0.98\tabularnewline
   ~~&   &   &   (1.12)&   (1.10)&   (1.16)&   (1.14)&   (1.13)&   (1.13)&   (1.12)&   (1.06)\tabularnewline
\shadeRow   ~~&   dHS Mix (approx.)&   &   1.10&   1.27&   1.06&   1.11&   1.14&   1.13&   1.08&   1.00\tabularnewline
\shadeRow   ~~&   &   &   (1.04)&   (1.09)&   (1.05)&   (1.06)&   (1.05)&   (1.04)&   (1.04)&   (0.98)\tabularnewline
   ~~&   dHS MS&   &   1.06&   1.11&   1.11&   1.14&   1.13&   1.11&   1.03&   0.92\tabularnewline
   ~~&   &   &   (1.07)&   (1.09)&   (1.12)&   (1.10)&   (1.09)&   (1.08)&   (1.07)&   (1.00)\tabularnewline
\shadeRow   ~~&   dHS MS (approx.)&   &   0.98&   1.01&   1.06&   1.08&   1.07&   1.02&   0.94&   0.81\tabularnewline
\shadeRow   ~~&   &   &   (0.99)&   (1.02)&   (1.05)&   (1.03)&   (1.01)&   (0.99)&   (0.97)&   (0.88)\tabularnewline
   ~~&   dHS svol-N&   &   1.07&   1.14&   1.15&   1.14&   1.14&   1.11&   1.04&   0.92\tabularnewline
   ~~&   &   &   (1.07)&   (1.09)&   (1.11)&   (1.09)&   (1.09)&   (1.08)&   (1.07)&   (0.99)\tabularnewline
\shadeRow   ~~&   dHS svol-N (approx.)&   &   0.92&   0.99&   0.98&   0.99&   0.99&   0.96&   0.89&   0.78\tabularnewline
\shadeRow   ~~&   &   &   (0.97)&   (1.00)&   (1.02)&   (1.00)&   (0.99)&   (0.97)&   (0.95)&   (0.87)\tabularnewline
   ~~&   dHS svol-Z&   &   1.11&   1.19&   1.21&   1.19&   1.18&   1.14&   1.07&   0.96\tabularnewline
   ~~&   &   &   (1.10)&   (1.11)&   (1.14)&   (1.12)&   (1.11)&   (1.11)&   (1.10)&   (1.03)\tabularnewline
\shadeRow   ~~&   dHS svol-Z (approx.)&   &   0.92&   1.03&   1.00&   0.99&   0.98&   0.94&   0.88&   0.78\tabularnewline
\shadeRow   ~~&   &   &   (0.97)&   (1.02)&   (1.02)&   (1.00)&   (0.99)&   (0.96)&   (0.94)&   (0.87)\tabularnewline
   ~~&   sHS&   &   1.09&   1.09&   1.15&   1.15&   1.15&   1.13&   1.08&   0.96\tabularnewline
   ~~&   &   &   (1.13)&   (1.10)&   (1.16)&   (1.15)&   (1.14)&   (1.14)&   (1.14)&   (1.07)\tabularnewline
\shadeRow   ~~&   sHS (approx.)&   &   0.91&   1.00&   0.97&   0.98&   0.97&   0.94&   0.88&   0.78\tabularnewline
\shadeRow   ~~&   &   &   (0.96)&   (1.01)&   (1.02)&   (1.00)&   (0.98)&   (0.96)&   (0.94)&   (0.86)\tabularnewline
\midrule
\multicolumn{11}{c}{\textbf{Nelson-Siegel TVP-VAR with the flexible specification for $\bm W$}}\tabularnewline
   ~~&   dHS Mix&   &   1.19&   1.25&   1.38&   1.36&   1.28&   1.20&   1.10&   0.99\tabularnewline
   ~~&   &   &   (1.33)&   (1.32)&   (1.43)&   (1.40)&   (1.35)&   (1.31)&   (1.29)&   (1.27)\tabularnewline
\shadeRow   ~~&   dHS Mix (approx.)&   &   \textbf{0.90}&   1.02&   0.98&   0.98&   0.97&   \textbf{0.93}&   \textbf{0.85}&   \textbf{0.75}\tabularnewline
\shadeRow   ~~&   &   &   (0.95)&   (1.01)&   (1.03)&   (1.00)&   (0.98)&   (\textbf{0.95})&   (\textbf{0.92})&   (\textbf{0.85})\tabularnewline
   ~~&   dHS MS&   &   1.00&   \textbf{0.93}&   1.13&   1.13&   1.10&   1.04&   0.95&   0.83\tabularnewline
   ~~&   &   &   (1.01)&   (0.99)&   (1.06)&   (1.05)&   (1.04)&   (1.01)&   (0.99)&   (0.91)\tabularnewline
\shadeRow   ~~&   dHS MS (approx.)&   &   1.06&   1.20&   1.20&   1.18&   1.16&   1.09&   0.98&   0.85\tabularnewline
\shadeRow   ~~&   &   &   (1.01)&   (1.05)&   (1.07)&   (1.06)&   (1.04)&   (1.02)&   (0.98)&   (0.90)\tabularnewline
   ~~&   dHS svol-N&   &   1.11&   1.26&   1.09&   1.11&   1.14&   1.14&   1.08&   1.02\tabularnewline
   ~~&   &   &   (1.11)&   (1.18)&   (1.14)&   (1.13)&   (1.13)&   (1.11)&   (1.10)&   (1.05)\tabularnewline
\shadeRow   ~~&   dHS svol-N (approx.)&   &   0.92&   1.01&   0.98&   0.99&   0.98&   0.95&   0.88&   0.78\tabularnewline
\shadeRow   ~~&   &   &   (0.97)&   (1.01)&   (1.02)&   (1.01)&   (0.99)&   (0.97)&   (0.95)&   (0.87)\tabularnewline
   ~~&   dHS svol-Z&   &   1.07&   1.20&   1.18&   1.17&   1.14&   1.09&   1.01&   0.92\tabularnewline
   ~~&   &   &   (1.23)&   (1.23)&   (1.29)&   (1.28)&   (1.26)&   (1.22)&   (1.20)&   (1.16)\tabularnewline
\shadeRow   ~~&   dHS svol-Z (approx.)&   &   0.93&   1.09&   1.07&   1.03&   1.00&   0.94&   0.86&   0.75\tabularnewline
\shadeRow   ~~&   &   &   (0.97)&   (1.03)&   (1.04)&   (1.01)&   (0.99)&   (0.97)&   (0.94)&   (0.86)\tabularnewline
   ~~&   sHS&   &   0.93&   1.11&   0.97&   0.98&   0.99&   0.96&   0.90&   0.80\tabularnewline
   ~~&   &   &   (0.98)&   (1.03)&   (1.02)&   (1.01)&   (1.00)&   (0.98)&   (0.97)&   (0.89)\tabularnewline
\shadeRow   ~~&   sHS (approx.)&   &   0.91&   1.01&   0.99&   0.98&   0.98&   0.94&   0.87&   0.77\tabularnewline
\shadeRow   ~~&   &   &   (0.96)&   (1.02)&   (1.03)&   (1.00)&   (0.98)&   (0.96)&   (0.94)&   (0.86)\tabularnewline
\midrule
\multicolumn{11}{c}{\textbf{Nelson-Siegel TVP-VAR with the conventional \cite{primiceri2005time} setup}}\tabularnewline
~~&   &   &   0.99&   1.05&   1.14&   1.09&   1.06&   1.01&   0.93&   0.82\tabularnewline
~~&   &   &   (1.02)&   (1.06)&   (1.12)&   (1.07)&   (1.03)&   (1.01)&   (0.99)&   (0.92)\tabularnewline
\bottomrule
\caption*{\scriptsize\textit{Notes}: This table displays the one-step ahead forecast performance for non-sparsified models. We focus on seven maturities ($1$y, $3$y, $5$y, $7$y, $10$y, $15$y, and $30$y) as our target variables and use a hold-out period from $2009$:$01$ to $2019$:$12$. Point forecast performance is measured by relative root mean square errors (RMSEs), while density forecast performance (shown in parentheses) by relative continuous ranked probability scores (CRPSs). We consider two different models in terms of the dimension of the (TVP-)VARs: a large model including all $30$ maturities ($M = 30$) and a small model specified as a three factor Nelson-Siegel model ($M = 3$). For the main TVP-VARs, we consider a flexible and a RW specification of $\bm W$, each with five different global-local shrinkage priors (four dynamic and one static). These TVP-VARs are estimated with two different algorithms: our proposed approximate approach and an exact algorithm. In addition, we consider the conventional TVP-VAR setup of \cite{primiceri2005time} for the Nelson-Siegel model and a set of VARs with constant coefficients. For the VARs with constant parameters, we adopt either a Minnesota or a horseshoe (HS) shrinkage prior. As overall benchmark model we choose a large VAR with constant parameters and a Minnesota prior. The red shaded rows correspond to the actual RMSE and CRPS values of this benchmark model, while the grey shaded rows correspond to models for which we use our approximate (but non-sparsified) MCMC algorithm. The best performing specification is in bold.}
\end{longtable}}

\clearpage

\section{Closing remarks}\label{sec:concl}
VARs modelled with many macroeconomic and financial data sets exhibit parameter change and structural breaks. Typically, most parameter change is found in the error covariance matrix. But there can be small amounts of time-variation in VAR coefficients where only some coefficients change and even they only change at points in time. The problem is how to uncover TVPs of this sort. Simply working with a model where all VAR coefficients change can lead to over-fitting and poor forecast performance. In light of this situation, one contribution of this paper lies in our development of  several dynamic horseshoe priors which are designed for picking up the kind of parameter change that often occurs in practice. In an application involving eurozone yield data our methods find small amounts of time variation in parameters. In a forecasting exercise we find that appropropriately modeling this time variation leads to forecast improvements. 

The second contribution of this paper lies in computation. The approximate MCMC algorithm developed in this paper is scalable in a manner that exact MCMC algorithms are not. Thus, we have developed an algorithm which can be used in the huge dimensional models that are increasingly being used by economists. 
Finally, we have developed an MCMC algorithm for common stochastic volatility specifications which is particularly well-suited for large $k$ applications such as the one considered in this paper. 

\newpage
\small{\setstretch{0.85}
\addcontentsline{toc}{section}{References}
\bibliographystyle{custom}
\bibliography{dhs}

\begin{thebibliography}{42}
\newcommand{\enquote}[1]{``#1''}
\providecommand{\natexlab}[1]{#1}

\bibitem[{Belmonte \emph{et~al.}(2014)Belmonte, Koop, and Korobilis}]{bkk}
\textsc{Belmonte M, Koop G, and Korobilis D} (2014), \enquote{Hierarchical
  shrinkage in time-varying coefficient models,} \emph{Journal of Forecasting}
  \textbf{33}(1), 80--94.

\bibitem[{Bhattacharya \emph{et~al.}(2016)Bhattacharya, Chakraborty, and
  Mallick}]{bhattacharya2016fast}
\textsc{Bhattacharya A, Chakraborty A, and Mallick BK} (2016), \enquote{Fast
  sampling with Gaussian scale mixture priors in high-dimensional regression,}
  \emph{Biometrika} \textbf{103}(4), 985--991.

\bibitem[{Bhattacharya \emph{et~al.}(2015)Bhattacharya, Pati, Pillai, and
  Dunson}]{bhattacharya2015dirichlet}
\textsc{Bhattacharya A, Pati D, Pillai NS, and Dunson DB} (2015),
  \enquote{Dirichlet--Laplace priors for optimal shrinkage,} \emph{Journal of
  the American Statistical Association} \textbf{110}(512), 1479--1490.

\bibitem[{Carriero \emph{et~al.}(2019)Carriero, Clark, and
  Marcellino}]{carriero2019large}
\textsc{Carriero A, Clark TE, and Marcellino M} (2019), \enquote{Large Bayesian
  vector autoregressions with stochastic volatility and non-conjugate priors,}
  \emph{Journal of Econometrics} \textbf{212}(1), 137--154.

\bibitem[{Carvalho \emph{et~al.}(2010)Carvalho, Polson, and
  Scott}]{carvalho2010horseshoe}
\textsc{Carvalho CM, Polson NG, and Scott JG} (2010), \enquote{The horseshoe
  estimator for sparse signals,} \emph{Biometrika} \textbf{97}(2), 465--480.

\bibitem[{Chan \emph{et~al.}(2020)Chan, Eisenstat, and
  Strachan}]{chan2020reducing}
\textsc{Chan JC, Eisenstat E, and Strachan RW} (2020), \enquote{Reducing the
  state space dimension in a large TVP-VAR,} \emph{Journal of Econometrics}
  \textbf{218}(1), 105--118.

\bibitem[{Chan and Jeliazkov(2009)}]{chan2009efficient}
\textsc{Chan JC, and Jeliazkov I} (2009), \enquote{Efficient simulation and
  integrated likelihood estimation in state space models,} \emph{International
  Journal of Mathematical Modelling and Numerical Optimisation}
  \textbf{1}(1-2), 101--120.

\bibitem[{Clark(2011)}]{clark2011}
\textsc{Clark TE} (2011), \enquote{Real-Time Density Forecasts From Bayesian
  Vector Autoregressions With Stochastic Volatility,} \emph{Journal of Business
  \& Economic Statistics} \textbf{29}(3), 327--341.

\bibitem[{Cogley \emph{et~al.}(2010)Cogley, Primiceri, and
  Sargent}]{cogley2010inflation}
\textsc{Cogley T, Primiceri GE, and Sargent TJ} (2010), \enquote{Inflation-gap
  persistence in the US,} \emph{American Economic Journal: Macroeconomics}
  \textbf{2}(1), 43--69.

\bibitem[{Diebold \emph{et~al.}(2006)Diebold, Rudebusch, and
  Aruoba}]{diebold2006macroeconomy}
\textsc{Diebold FX, Rudebusch GD, and Aruoba SB} (2006), \enquote{The
  macroeconomy and the yield curve: a dynamic latent factor approach,}
  \emph{Journal of Econometrics} \textbf{131}(1-2), 309--338.

\bibitem[{Eisenstat \emph{et~al.}(2016)Eisenstat, Chan, and
  Strachan}]{eisenstat2016stochastic}
\textsc{Eisenstat E, Chan JC, and Strachan RW} (2016), \enquote{Stochastic
  model specification search for time-varying parameter VARs,}
  \emph{Econometric Reviews} \textbf{35}(8-10), 1638--1665.

\bibitem[{Fischer \emph{et~al.}(2023)Fischer, Hauzenberger, Huber, and
  Pfarrhofer}]{fischer2023general}
\textsc{Fischer MM, Hauzenberger N, Huber F, and Pfarrhofer M} (2023),
  \enquote{General Bayesian time-varying parameter vector autoregressions for
  modeling government bond yields,} \emph{Journal of Applied Econometrics}
  \textbf{38}(1), 69--87.

\bibitem[{Gneiting and Raftery(2007)}]{gneiting2007strictly}
\textsc{Gneiting T, and Raftery AE} (2007), \enquote{Strictly proper scoring
  rules, prediction, and estimation,} \emph{Journal of the American statistical
  Association} \textbf{102}(477), 359--378.

\bibitem[{Griffin and Brown(2010)}]{griffin2010inference}
\textsc{Griffin J, and Brown P} (2010), \enquote{Inference with normal-gamma
  prior distributions in regression problems,} \emph{Bayesian Analysis}
  \textbf{5}(1), 171--188.

\bibitem[{Hahn and Carvalho(2015)}]{hahncarvalho2015dss}
\textsc{Hahn PR, and Carvalho CM} (2015), \enquote{Decoupling Shrinkage and
  Selection in Bayesian Linear Models: A Posterior Summary Perspective,}
  \emph{Journal of the American Statistical Association} \textbf{110}(509),
  435--448.

\bibitem[{Hauzenberger(2021)}]{hauzenberger2021flexible}
\textsc{Hauzenberger N} (2021), \enquote{Flexible mixture priors for large
  time-varying parameter models,} \emph{Econometrics and Statistics}
  \textbf{20}, 87--108.

\bibitem[{Hauzenberger \emph{et~al.}(2022)Hauzenberger, Huber, Koop, and
  Onorante}]{hauzenberger2022fast}
\textsc{Hauzenberger N, Huber F, Koop G, and Onorante L} (2022), \enquote{Fast
  and flexible Bayesian inference in time-varying parameter regression models,}
  \emph{Journal of Business \& Economic Statistics} \textbf{40}(4), 1904--1918.

\bibitem[{Huber \emph{et~al.}(2021)Huber, Koop, and Onorante}]{hko2020}
\textsc{Huber F, Koop G, and Onorante L} (2021), \enquote{Inducing sparsity and
  shrinkage in time-varying parameter models,} \emph{Journal of Business \&
  Economic Statistics} \textbf{39}(3), 669--683.

\bibitem[{Huber \emph{et~al.}(2020)Huber, Koop, and Pfarrhofer}]{hkp2020}
\textsc{Huber F, Koop G, and Pfarrhofer M} (2020), \enquote{Bayesian Inference
  in High-Dimensional Time-varying Parameter Models using Integrated Rotated
  Gaussian Approximations,} \emph{arXiv preprint arXiv:2002.10274} .

\bibitem[{Ishwaran and Rao(2005)}]{ishwaran2005spike}
\textsc{Ishwaran H, and Rao JS} (2005), \enquote{Spike and slab variable
  selection: Frequentist and Bayesian strategies,} \emph{The Annals of
  Statistics} \textbf{33}(2), 730--773.

\bibitem[{Jacquier \emph{et~al.}(1995)Jacquier, Polson, and Rossi}]{JPR1995}
\textsc{Jacquier E, Polson N, and Rossi P} (1995), \enquote{Models and Priors
  for Multivariate Stochastic Volatility Models,} Technical report, Technical
  Report, University of Chicago, Graduate School of Business.

\bibitem[{Johndrow \emph{et~al.}(2020)Johndrow, Orenstein, and
  Bhattacharya}]{johndrow2020scalable}
\textsc{Johndrow J, Orenstein P, and Bhattacharya A} (2020), \enquote{Scalable
  Approximate MCMC Algorithms for the Horseshoe Prior,} \emph{Journal of
  Machine Learning Research} \textbf{21}(73), 1--61.

\bibitem[{Johndrow \emph{et~al.}(2017)Johndrow, Orenstein, and
  Bhattacharya}]{johndrow2017bayes}
\textsc{Johndrow JE, Orenstein P, and Bhattacharya A} (2017), \enquote{Bayes
  shrinkage at GWAS scale: Convergence and approximation theory of a scalable
  MCMC algorithm for the horseshoe prior,} \emph{arXiv preprint
  arXiv:1705.00841} .

\bibitem[{Kalli and Griffin(2014)}]{kg2014}
\textsc{Kalli M, and Griffin J} (2014), \enquote{Time-varying sparsity in
  dynamic regression models,} \emph{Journal of Econometrics} \textbf{178}(2),
  779 -- 793.

\bibitem[{Kalli and Griffin(2019)}]{kalli2019bayesian}
---{}---{}--- (2019), \enquote{Bayesian nonparametric time varying vector
  autoregressive models,} \emph{Journal of Business \& Economic Statistics} .

\bibitem[{Kastner and Fr{\"u}hwirth-Schnatter(2014)}]{kastner2014ancillarity}
\textsc{Kastner G, and Fr{\"u}hwirth-Schnatter S} (2014),
  \enquote{Ancillarity-sufficiency interweaving strategy (ASIS) for boosting
  MCMC estimation of stochastic volatility models,} \emph{Computational
  Statistics \& Data Analysis} \textbf{76}, 408--423.

\bibitem[{Kastner and Huber(2020)}]{kastner2020sparse}
\textsc{Kastner G, and Huber F} (2020), \enquote{Sparse Bayesian vector
  autoregressions in huge dimensions,} \emph{Journal of Forecasting}
  \textbf{39}(7), 1142--1165.

\bibitem[{Kim and Nelson(1999{\natexlab{a}})}]{kim1999has}
\textsc{Kim CJ, and Nelson CR} (1999{\natexlab{a}}), \enquote{Has the US
  economy become more stable? A Bayesian approach based on a Markov-switching
  model of the business cycle,} \emph{Review of Economics and Statistics}
  \textbf{81}(4), 608--616.

\bibitem[{Kim and Nelson(1999{\natexlab{b}})}]{kim1999state}
---{}---{}--- (1999{\natexlab{b}}), \enquote{State-space models with regime
  switching: classical and Gibbs-sampling approaches with applications,}
  \emph{MIT Press Books} \textbf{1}.

\bibitem[{Kim \emph{et~al.}(1998)Kim, Shephard, and Chib}]{kim1998stochastic}
\textsc{Kim S, Shephard N, and Chib S} (1998), \enquote{Stochastic volatility:
  likelihood inference and comparison with ARCH models,} \emph{The Review of
  Economic Studies} \textbf{65}(3), 361--393.

\bibitem[{Knaus \emph{et~al.}(2021)Knaus, Bitto-Nemling, Cadonna, and
  Fr{\"u}hwirth-Schnatter}]{bitto2019shrinkage}
\textsc{Knaus P, Bitto-Nemling A, Cadonna A, and Fr{\"u}hwirth-Schnatter S}
  (2021), \enquote{Shrinkage in the time-varying parameter model framework
  using the R package shrinkTVP,} \emph{Journal of Statistical Software}
  \textbf{100}(13), 1--32.

\bibitem[{Korobilis(2021)}]{korobilis2019high}
\textsc{Korobilis D} (2021), \enquote{High-dimensional macroeconomic
  forecasting using message passing algorithms,} \emph{Journal of Business \&
  Economic Statistics} \textbf{39}(2), 493--504.

\bibitem[{Korobilis(2022)}]{korobilis2022new}
---{}---{}--- (2022), \enquote{A new algorithm for structural restrictions in
  Bayesian vector autoregressions,} \emph{European Economic Review}
  \textbf{148}, 104241.

\bibitem[{Kowal \emph{et~al.}(2019)Kowal, Matteson, and
  Ruppert}]{kowal2019dynamic}
\textsc{Kowal DR, Matteson DS, and Ruppert D} (2019), \enquote{Dynamic
  shrinkage processes,} \emph{Journal of the Royal Statistical Society: Series
  B (Statistical Methodology)} \textbf{81}(4), 781--804.

\bibitem[{Makalic and Schmidt(2015)}]{makalic2015simple}
\textsc{Makalic E, and Schmidt DF} (2015), \enquote{A simple sampler for the
  horseshoe estimator,} \emph{IEEE Signal Processing Letters} \textbf{23}(1),
  179--182.

\bibitem[{McCausland \emph{et~al.}(2011)McCausland, Miller, and
  Pelletier}]{MCCAUSLAND2011199}
\textsc{McCausland WJ, Miller S, and Pelletier D} (2011), \enquote{Simulation
  smoothing for state--space models: A computational efficiency analysis,}
  \emph{Computational Statistics \& Data Analysis} \textbf{55}(1), 199--212.

\bibitem[{Nelson and Siegel(1987)}]{nelson1987parsimonious}
\textsc{Nelson CR, and Siegel AF} (1987), \enquote{Parsimonious modeling of
  yield curves,} \emph{Journal of Business} 473--489.

\bibitem[{Park and Casella(2008)}]{park2008bayesian}
\textsc{Park T, and Casella G} (2008), \enquote{The Bayesian Lasso,}
  \emph{Journal of the American Statistical Association} \textbf{103}(482),
  681--686.

\bibitem[{Petrova(2019)}]{petrova2019quasi}
\textsc{Petrova K} (2019), \enquote{A quasi-Bayesian local likelihood approach
  to time varying parameter VAR models,} \emph{Journal of Econometrics}
  \textbf{212}(1), 286--306.

\bibitem[{Primiceri(2005)}]{primiceri2005time}
\textsc{Primiceri G} (2005), \enquote{Time varying structural autoregressions
  and monetary policy,} \emph{Oxford University Press} \textbf{72}(3),
  821--852.

\bibitem[{Puelz \emph{et~al.}(2020)Puelz, Hahn, and
  Carvalho}]{puelz2020portfolio}
\textsc{Puelz D, Hahn PR, and Carvalho CM} (2020), \enquote{Portfolio selection
  for individual passive investing,} \emph{Applied Stochastic Models in
  Business and Industry} \textbf{36}(1), 124--142.

\bibitem[{Ray and Bhattacharya(2018)}]{bhattacharya2018signal}
\textsc{Ray P, and Bhattacharya A} (2018), \enquote{Signal Adaptive Variable
  Selector for the Horseshoe Prior,} \emph{arXiv preprint arXiv:1810.09004} .

\end{thebibliography}

\clearpage
\begin{appendices}
\setcounter{equation}{0}
\renewcommand\theequation{A.\arabic{equation}}
\section{Details of the MCMC Algorithm}
\doublespacing
\subsection{Sampling the Log-Volatilities}\label{sec:logvola}

We assume a stochastic volatility process of the following form for $h_t = log(\sigma^2_t)$:
\begin{equation*}
h_t = \mu_h + \rho_h (h_{t-1}-\mu_h) + \sigma_h v_t, \quad v_t \sim \mathcal{N}(0, 1), \quad h_0 \sim \mathcal{N}\left(\mu, \frac{\sigma_h^2}{1-\rho_h^2}\right).
\end{equation*}
Following \cite{kastner2014ancillarity} we make the prior assumptions that $\mu_h \sim \mathcal{N}(0, 10)$, $\frac{\rho_h+1}{2}\sim \mathcal{B}(5, 1.5)$ and $\sigma_h^2 \sim \mathcal{G}(1/2, 1/2)$ where $\mathcal{B}$ and $\mathcal{G}$ denote the Beta and Gamma distributions, respectively. We use the algorithm of \cite{kastner2014ancillarity} to take draws of $h_t$.

\subsection{Sampling the Time-Invariant Regression Coefficients}
Most of the conditional posterior distributions take a simple and well-known form. Here we briefly summarize these and provide some information on the relevant literature.

The time-invariant coefficients $\bm \alpha$ follow a $K$-dimensional multivariate Gaussian posterior given by
\begin{align*}
\bm \alpha | \bullet &\sim \mathcal{N}(\overline{\bm \alpha}, \overline{\bm V}_\alpha), \\
\overline{\bm V}_\alpha &=\left (\tilde{\bm X} ' \tilde{\bm X} + \bm D^{-1}_\alpha\right)^{-1},\\
  \overline{\bm \alpha} &= \overline{\bm V}_\alpha \tilde{\bm X} \hat{\bm y}, \label{eq: postalpha}
\end{align*}
with $\tilde{\bm X} = \bm L^{-1} \bm X$, $\hat{\bm y} = \bm L^{-1}(\bm y - \bm W \bm \beta)$ and $\bm D_\alpha = \tau_\alpha~ \text{diag}(\psi^2_1, \dots, \psi^2_K)$ denoting a $K\times K$-dimensional prior variance-covariance matrix with $\psi_j~(j=1,\dots, K)$ and $\sqrt{\tau}_\alpha$ following a half-Cauchy distribution, respectively.

\subsection{Sampling the horseshoe Prior on the  Constant and the  Time-varying Parameters}
 \cite{makalic2015simple} show that one can simulate from the posterior distribution of $\psi_j$ using standard distributions only. This is achieved by introducing additional auxiliary quantities $\varrho_j ~(j=1, \dots, K)$. Using these, the posterior of $\psi_j$ follows an inverted Gamma distribution:
\begin{equation*}
\psi^2_j | \bullet \sim \mathcal{G}^{-1}\left(1, \frac{1}{\varrho_j} + \frac{\alpha_j^2}{2 \tau_\alpha}\right)
\end{equation*}
where $\alpha_j$ denotes the $j^{th}$ element of $\bm \alpha$. The posterior of $\varrho_j$ is also inverse Gamma distributed with $\varrho_j|\bullet \sim \mathcal{G}^{-1}(1, 1+\psi^{-2}_j)$.  

For the global shrinkage parameter, we introduce yet another auxiliary quantity $\varpi_\alpha$. This enables us to derive a conditional posterior for $\tau_\alpha$ which is also inverse Gamma distributed:
\begin{equation*}
\tau_\alpha | \bullet \sim \mathcal{G}^{-1}\left(\frac{K+1}{2}, \frac{1}{\varpi_\alpha} + \sum_{j=1}^K \frac{\alpha_j^2 }{2 \psi_j^2}\right)
\end{equation*}
and the posterior of $\varpi_\alpha$ being given by:
\begin{equation*}
\varpi_\alpha | \bullet \sim \mathcal{G}^{-1}(1, 1+\tau_\alpha^{-1}).
\end{equation*}

The local shrinkage parameters $\phi_{jt}$ can be simulated conditionally on $\tau$ and $\{\lambda_t\}_{t=1}^T$ similarly to the $\psi_j$'s. Specificially, the posterior distribution of $\phi^2_{jt}$ follows an inverse Gamma:
\begin{equation*}
\phi^2_{jt}|\bullet \sim \mathcal{G}^{-1}\left(1, \frac{1}{\vartheta_{jt}} + \frac{\beta_{jt}^2}{2 \tau \lambda_t}\right)
\end{equation*}
with $\vartheta_{jt}$ denoting yet another scaling parameter that follows a inverse Gamma posterior distribution: $\vartheta_{jt}|\bullet \sim \mathcal{G}^{-1}(1, 1 + \phi_{jt}^{-2})$.

If we do not assume $\lambda_t$ to evolve according to an AR(1) process, we sample the global shrinkage parameter $\tau$ similar to $\tau_\alpha$. The conditional posterior of $\tau$ also follows an inverse Gamma: 
\begin{equation*}
\tau | \bullet \sim \mathcal{G}^{-1}\left(\frac{k+1}{2}, \frac{1}{\varpi} + \sum_{t=1}^T \sum_{j=1}^K \frac{\beta_{jt}^2}{2 \lambda_t \phi_{jt}^2}\right)
\end{equation*}
with the posterior of the auxiliary variable $\varpi$ given by:
\begin{equation*}
\varpi | \bullet \sim \mathcal{G}^{-1}(1, 1+\tau^{-1}).
\end{equation*}

\subsection{Sampling the Dynamic Shrinkage Parameters}

As stated in Sub-section \ref{sec: shrinkagePRIOR}, the full history of $\lambda_t$  in the case that it follows a mixture or Markov switching specification can be easily obtained through standard techniques.  More precisely, if $d_t$ in  (\ref{eq: mixlambda}) follows a Markov switching model, we adopt the algorithm discussed in, e.g., \cite{kim1999state, kim1999has}. The  posterior of the transition probabilities is Beta distributed:
\begin{equation*}
p_{ii}|\bullet \sim \mathcal{B}(a_{i,MS} +T_{i0}, b_{i,MS} + T_{i1}),
\end{equation*}
whereby $T_{ij}$ denotes the number of times a transition from state $i$ to $j$ has been observed in the full history of $d_t$.

In the case of the mixture model, the posterior distribution of $d_t$ follows a Bernoulli distribution for each $t$:
\begin{equation*}
Prob(d_t = 1 | \bullet)  = {Ber}(\overline{p}_t)
\end{equation*} 
with $\overline{p}_t$ given by:
\begin{equation*}
\overline{p}_t = \frac{\kappa_1^{- K/2} \exp{ \left( -\frac{\sum_{j=1}^K \hat{\beta}_{jt}}{2 \kappa_1^2}\right)} \times \underline{p}}{\kappa_1^{- K/2} \exp{ \left( -\frac{\sum_{j=1}^K \hat{\beta}_{jt}}{2 \kappa_1^2}\right)} \times \underline{p} + \kappa_0^{- K/2} \exp{ \left( -\frac{\sum_{j=1}^K \hat{\beta}_{jt}}{2 \kappa_0^2}\right)} \times (1-\underline{p})}.
\end{equation*}
and the posterior of $\underline{p}$ follows a Beta distribution $\underline{p} | \bullet \sim \mathcal{B}\left(\sum_{t=1}^T d_t + a_{Mix}, 1 - \sum_{t=1}^T d_t + b_{Mix}\right)$.

Finally, in the case that $\lambda_t$ evolves according to an AR(1) process with Gaussian shocks, we use precisely the same algorithm as  \cite{kastner2014ancillarity} for simulating $\mu$ and $\rho$. In the case that we use Z-distributed shocks, the algorithm proposed in \cite{kowal2019dynamic} is adopted. This implies that we use Polya-Gamma (PG) auxiliary random variables to approximate the Z-distribution using a scale-mixture of Gaussians. Essentially, the main implication is that conditional on the $T$ PG random variates, the parameters of the state evolution equation can be estimated similarly to the Gaussian case after normalizing everything by rendering the AR(1) conditionally homoscedastic. For more details, see \cite{kowal2019dynamic}.

\section{Higher-order forecast performance}
\setcounter{table}{0}
\renewcommand\thetable{B.\arabic{table}}
\clearpage
{\tiny
\begin{longtable}{llcllllllll}
\caption{One-quarter ahead forecast performance for EA central government bond yields at different maturities using non-sparsified models. \label{tab:non-sps3}}\tabularnewline
\toprule
\multicolumn{1}{l}{\bfseries }&\multicolumn{1}{c}{\bfseries Specification}&\multicolumn{1}{c}{\bfseries }&\multicolumn{8}{c}{\bfseries One-quarter-ahead}\tabularnewline
\cmidrule{4-11}
\multicolumn{1}{l}{}&\multicolumn{1}{c}{}&\multicolumn{1}{c}{}&\multicolumn{1}{c}{Avg.}&\multicolumn{1}{c}{1y}&\multicolumn{1}{c}{3y}&\multicolumn{1}{c}{5y}&\multicolumn{1}{c}{7y}&\multicolumn{1}{c}{10y}&\multicolumn{1}{c}{15y}&\multicolumn{1}{c}{30y}\tabularnewline
\midrule
\endfirsthead
\multicolumn{11}{r}{\footnotesize{\textit{Table \ref{tab:non-sps3} continued}}}\\
\toprule
\multicolumn{1}{l}{\bfseries }&\multicolumn{1}{c}{\bfseries Specification}&\multicolumn{1}{c}{\bfseries }&\multicolumn{8}{c}{\bfseries One-quarter-ahead}\tabularnewline
\cmidrule{4-11}
\multicolumn{1}{l}{}&\multicolumn{1}{c}{}&\multicolumn{1}{c}{}&\multicolumn{1}{c}{Avg.}&\multicolumn{1}{c}{1y}&\multicolumn{1}{c}{3y}&\multicolumn{1}{c}{5y}&\multicolumn{1}{c}{7y}&\multicolumn{1}{c}{10y}&\multicolumn{1}{c}{15y}&\multicolumn{1}{c}{30y}\tabularnewline
\midrule
\endhead
\multicolumn{11}{c}{\textbf{VAR with constant coefficients}}\tabularnewline
\shadeBench   ~~&   Large VAR with MIN prior&   &   0.97&   0.74&   0.83&   0.95&   1.03&   1.08&   1.10&   0.99\tabularnewline
\shadeBench   ~~&   &   &   (0.53)&   (0.36)&   (0.45)&   (0.53)&   (0.58)&   (0.61)&   (0.62)&   (0.56)\tabularnewline
   ~~&   Large VAR with HS prior&   &   0.95&   0.92&   0.94&   0.94&   0.94&   0.94&   0.96&   1.00\tabularnewline
   ~~&   &   &   (0.95)&   (0.95)&   (0.96)&   (0.96)&   (0.95)&   (0.95)&   (0.95)&   (0.96)\tabularnewline
   ~~&   Nelson-Siegel VAR with HS prior&   &   0.95&   1.02&   0.97&   0.94&   0.92&   0.92&   0.94&   0.98\tabularnewline
   ~~&   &   &   (0.95)&   (1.03)&   (1.01)&   (0.96)&   (0.93)&   (0.92)&   (0.93)&   (0.94)\tabularnewline
   ~~&   Nelson-Siegel VAR with MIN prior&   &   0.96&   1.02&   1.00&   0.95&   0.93&   0.93&   0.95&   0.99\tabularnewline
   ~~&   &   &   (0.97)&   (1.05)&   (1.03)&   (0.97)&   (0.95)&   (0.94)&   (0.94)&   (0.95)\tabularnewline
\midrule
\multicolumn{11}{c}{\textbf{Large TVP-VAR with random walk specification for $\bm W$}}\tabularnewline
   ~~&   dHS Mix&   &   1.06&   0.94&   0.96&   0.99&   1.01&   1.04&   1.09&   1.25\tabularnewline
   ~~&   &   &   (1.01)&   (0.96)&   (0.97)&   (0.99)&   (1.00)&   (1.02)&   (1.03)&   (1.07)\tabularnewline
\shadeRow   ~~&   dHS Mix (approx.)&   &   \textbf{0.94}&   \textbf{0.91}&   0.94&   0.93&   0.91&   \textbf{0.91}&   0.94&   1.00\tabularnewline
\shadeRow   ~~&   &   &   (0.96)&   (0.95)&   (\textbf{0.94})&   (\textbf{0.93})&   (0.93)&   (0.94)&   (0.97)&   (1.09)\tabularnewline
   ~~&   dHS MS&   &   0.97&   0.93&   0.95&   0.94&   0.93&   0.95&   0.99&   1.08\tabularnewline
   ~~&   &   &   (0.97)&   (0.95)&   (0.97)&   (0.96)&   (0.96)&   (0.96)&   (0.98)&   (1.01)\tabularnewline
\shadeRow   ~~&   dHS MS (approx.)&   &   0.94&   0.95&   0.93&   0.92&   0.92&   0.93&   0.94&   0.99\tabularnewline
\shadeRow   ~~&   &   &   (0.96)&   (0.95)&   (0.95)&   (0.95)&   (0.95)&   (0.95)&   (0.96)&   (0.99)\tabularnewline
   ~~&   dHS svol-N&   &   0.95&   0.92&   0.94&   0.94&   0.93&   0.94&   0.95&   1.01\tabularnewline
   ~~&   &   &   (0.97)&   (0.95)&   (0.96)&   (0.96)&   (0.95)&   (0.96)&   (0.97)&   (1.04)\tabularnewline
\shadeRow   ~~&   dHS svol-N (approx.)&   &   0.95&   0.94&   0.94&   0.94&   0.93&   0.94&   0.95&   1.00\tabularnewline
\shadeRow   ~~&   &   &   (1.10)&   (1.16)&   (1.13)&   (1.10)&   (1.08)&   (1.07)&   (1.07)&   (1.11)\tabularnewline
   ~~&   dHS svol-Z&   &   0.95&   0.93&   0.94&   0.93&   0.93&   0.94&   0.96&   1.01\tabularnewline
   ~~&   &   &   (1.09)&   (1.15)&   (1.12)&   (1.09)&   (1.07)&   (1.06)&   (1.07)&   (1.10)\tabularnewline
\shadeRow   ~~&   dHS svol-Z (approx.)&   &   0.96&   0.93&   0.96&   0.94&   0.93&   0.94&   0.96&   1.02\tabularnewline
\shadeRow   ~~&   &   &   (1.03)&   (0.95)&   (0.97)&   (0.97)&   (0.96)&   (1.08)&   (1.08)&   (1.13)\tabularnewline
   ~~&   sHS&   &   1.00&   0.97&   0.96&   0.96&   0.96&   0.99&   1.02&   1.07\tabularnewline
   ~~&   &   &   (1.13)&   (1.18)&   (1.14)&   (1.12)&   (1.11)&   (1.12)&   (1.12)&   (1.16)\tabularnewline
\shadeRow   ~~&   sHS (approx.)&   &   0.94&   0.93&   0.93&   0.93&   0.92&   0.92&   0.94&   0.99\tabularnewline
\shadeRow   ~~&   &   &   (1.02)&   (0.96)&   (0.96)&   (0.96)&   (0.95)&   (1.07)&   (1.08)&   (1.12)\tabularnewline
\midrule
\multicolumn{11}{c}{\textbf{Large TVP-VAR with the flexible specification for $\bm W$}}\tabularnewline
   ~~&   dHS Mix&   &   1.11&   0.96&   0.93&   0.95&   1.01&   1.11&   1.22&   1.36\tabularnewline
   ~~&   &   &   (1.10)&   (0.98)&   (0.99)&   (1.03)&   (1.07)&   (1.13)&   (1.18)&   (1.25)\tabularnewline
\shadeRow   ~~&   dHS Mix (approx.)&   &   0.99&   0.97&   0.99&   0.98&   0.96&   0.95&   0.97&   1.11\tabularnewline
\shadeRow   ~~&   &   &   (1.52)&   (1.06)&   (1.20)&   (1.26)&   (1.27)&   (1.37)&   (1.60)&   (2.68)\tabularnewline
   ~~&   dHS MS&   &   2.11&   0.95&   1.66&   2.06&   2.10&   2.22&   2.30&   2.49\tabularnewline
   ~~&   &   &   (2.58)&   (1.45)&   (2.02)&   (2.37)&   (2.53)&   (2.71)&   (2.90)&   (3.50)\tabularnewline
\shadeRow   ~~&   dHS MS (approx.)&   &   0.96&   0.94&   0.96&   0.95&   0.94&   0.95&   0.97&   1.01\tabularnewline
\shadeRow   ~~&   &   &   (1.26)&   (1.39)&   (1.31)&   (1.25)&   (1.22)&   (1.21)&   (1.21)&   (1.29)\tabularnewline
   ~~&   dHS svol-N&   &   0.96&   0.93&   0.93&   0.93&   0.93&   0.94&   0.98&   1.03\tabularnewline
   ~~&   &   &   (0.96)&   (0.95)&   (0.95)&   (0.95)&   (0.95)&   (0.95)&   (0.96)&   (0.97)\tabularnewline
\shadeRow   ~~&   dHS svol-N (approx.)&   &   0.94&   0.91&   \textbf{0.92}&   \textbf{0.92}&   0.92&   0.93&   0.95&   0.99\tabularnewline
\shadeRow   ~~&   &   &   (1.23)&   (1.36)&   (1.28)&   (1.23)&   (1.20)&   (1.19)&   (1.19)&   (1.22)\tabularnewline
   ~~&   dHS svol-Z&   &   0.98&   0.91&   0.93&   0.94&   0.94&   0.97&   1.01&   1.09\tabularnewline
   ~~&   &   &   (0.98)&   (0.95)&   (0.96)&   (0.97)&   (0.96)&   (0.98)&   (1.01)&   (1.03)\tabularnewline
\shadeRow   ~~&   dHS svol-Z (approx.)&   &   0.95&   0.97&   0.96&   0.95&   0.93&   0.93&   0.95&   0.99\tabularnewline
\shadeRow   ~~&   &   &   (0.95)&   (0.96)&   (0.96)&   (0.96)&   (0.95)&   (0.95)&   (0.95)&   (0.95)\tabularnewline
   ~~&   sHS&   &   0.95&   0.93&   0.95&   0.94&   0.93&   0.93&   0.95&   0.99\tabularnewline
   ~~&   &   &   (0.95)&   (0.95)&   (0.96)&   (0.95)&   (0.95)&   (0.94)&   (0.95)&   (0.94)\tabularnewline
\shadeRow   ~~&   sHS (approx.)&   &   0.94&   0.93&   0.94&   0.93&   0.92&   0.92&   0.94&   0.97\tabularnewline
\shadeRow   ~~&   &   &   (0.95)&   (\textbf{0.94})&   (0.96)&   (0.96)&   (0.95)&   (0.95)&   (0.94)&   (0.95)\tabularnewline
\midrule
\multicolumn{11}{r}{}\tabularnewline
\multicolumn{11}{r}{}\tabularnewline
\multicolumn{11}{r}{}\tabularnewline
\multicolumn{11}{c}{\textbf{Nelson-Siegel TVP-VAR with random walk specification for $\bm W$}}\tabularnewline
   ~~&   dHS Mix&   &   1.02&   1.02&   1.07&   1.03&   1.00&   0.99&   1.01&   1.05\tabularnewline
   ~~&   &   &   (1.08)&   (1.13)&   (1.17)&   (1.10)&   (1.05)&   (1.03)&   (1.03)&   (1.08)\tabularnewline
\shadeRow   ~~&   dHS Mix (approx.)&   &   1.37&   1.93&   1.47&   1.34&   1.27&   1.24&   1.24&   1.33\tabularnewline
\shadeRow   ~~&   &   &   (1.12)&   (1.35)&   (1.21)&   (1.12)&   (1.08)&   (1.06)&   (1.07)&   (1.10)\tabularnewline
   ~~&   dHS MS&   &   0.98&   0.96&   1.00&   0.97&   0.96&   0.96&   0.98&   1.04\tabularnewline
   ~~&   &   &   (1.05)&   (1.09)&   (1.12)&   (1.06)&   (1.02)&   (1.01)&   (1.02)&   (1.06)\tabularnewline
\shadeRow   ~~&   dHS MS (approx.)&   &   0.94&   0.95&   0.98&   0.94&   0.92&   0.92&   0.94&   0.98\tabularnewline
\shadeRow   ~~&   &   &   (0.96)&   (1.01)&   (1.02)&   (0.97)&   (0.94)&   (0.94)&   (0.94)&   (0.95)\tabularnewline
   ~~&   dHS svol-N&   &   0.98&   0.98&   1.02&   0.99&   0.97&   0.96&   0.97&   1.02\tabularnewline
   ~~&   &   &   (1.02)&   (1.08)&   (1.10)&   (1.03)&   (1.00)&   (0.98)&   (0.99)&   (1.02)\tabularnewline
\shadeRow   ~~&   dHS svol-N (approx.)&   &   0.97&   1.23&   0.96&   0.94&   0.92&   0.93&   0.95&   0.98\tabularnewline
\shadeRow   ~~&   &   &   (0.96)&   (1.08)&   (1.01)&   (0.96)&   (0.94)&   (0.93)&   (0.94)&   (0.94)\tabularnewline
   ~~&   dHS svol-Z&   &   1.00&   0.97&   1.04&   1.00&   0.98&   0.98&   0.99&   1.03\tabularnewline
   ~~&   &   &   (1.04)&   (1.10)&   (1.13)&   (1.06)&   (1.02)&   (1.00)&   (1.01)&   (1.05)\tabularnewline
\shadeRow   ~~&   dHS svol-Z (approx.)&   &   1.00&   1.53&   0.98&   0.94&   0.92&   0.92&   0.94&   0.97\tabularnewline
\shadeRow   ~~&   &   &   (0.96)&   (1.15)&   (1.01)&   (0.96)&   (0.93)&   (0.92)&   (0.93)&   (0.93)\tabularnewline
   ~~&   sHS&   &   1.28&   1.54&   1.15&   1.17&   1.22&   1.28&   1.28&   1.36\tabularnewline
   ~~&   &   &   (1.21)&   (1.32)&   (1.27)&   (1.21)&   (1.17)&   (1.16)&   (1.18)&   (1.24)\tabularnewline
\shadeRow   ~~&   sHS (approx.)&   &   0.98&   1.04&   1.09&   1.05&   0.97&   0.93&   \textbf{0.92}&   \textbf{0.96}\tabularnewline
\shadeRow   ~~&   &   &   (0.97)&   (1.05)&   (1.05)&   (1.00)&   (0.96)&   (0.93)&   (0.93)&   (0.93)\tabularnewline
\midrule
\multicolumn{11}{c}{\textbf{Nelson-Siegel TVP-VAR with the flexible specification for $\bm W$}}\tabularnewline
   ~~&   dHS Mix&   &   1.90&   2.10&   2.49&   1.89&   1.64&   1.59&   1.70&   2.15\tabularnewline
   ~~&   &   &   (2.13)&   (2.24)&   (2.55)&   (2.23)&   (2.01)&   (1.92)&   (1.94)&   (2.15)\tabularnewline
\shadeRow   ~~&   dHS Mix (approx.)&   &   1.22&   1.51&   1.42&   1.11&   1.26&   1.19&   1.12&   1.06\tabularnewline
\shadeRow   ~~&   &   &   (1.06)&   (1.18)&   (1.16)&   (1.09)&   (1.06)&   (1.03)&   (1.01)&   (0.97)\tabularnewline
   ~~&   dHS MS&   &   1.03&   1.04&   1.28&   1.10&   1.00&   0.96&   0.95&   0.99\tabularnewline
   ~~&   &   &   (1.00)&   (1.05)&   (1.09)&   (1.02)&   (0.98)&   (0.97)&   (0.97)&   (0.98)\tabularnewline
\shadeRow   ~~&   dHS MS (approx.)&   &   0.96&   0.99&   1.07&   0.97&   0.92&   0.92&   0.93&   0.97\tabularnewline
\shadeRow   ~~&   &   &   (0.98)&   (1.07)&   (1.06)&   (0.99)&   (0.96)&   (0.95)&   (0.96)&   (0.96)\tabularnewline
   ~~&   dHS svol-N&   &   1.37&   1.23&   1.63&   1.44&   1.33&   1.30&   1.32&   1.37\tabularnewline
   ~~&   &   &   (1.33)&   (1.36)&   (1.51)&   (1.38)&   (1.30)&   (1.26)&   (1.26)&   (1.29)\tabularnewline
\shadeRow   ~~&   dHS svol-N (approx.)&   &   0.94&   0.97&   0.97&   0.93&   0.91&   0.91&   0.93&   0.97\tabularnewline
\shadeRow   ~~&   &   &   (\textbf{0.95})&   (1.01)&   (1.00)&   (0.95)&   (\textbf{0.92})&   (\textbf{0.92})&   (0.93)&   (0.94)\tabularnewline
   ~~&   dHS svol-Z&   &   1.61&   1.50&   2.00&   1.70&   1.52&   1.46&   1.49&   1.68\tabularnewline
   ~~&   &   &   (1.73)&   (1.70)&   (2.04)&   (1.83)&   (1.68)&   (1.61)&   (1.62)&   (1.75)\tabularnewline
\shadeRow   ~~&   dHS svol-Z (approx.)&   &   0.94&   1.00&   0.95&   0.92&   \textbf{0.91}&   0.91&   0.94&   0.98\tabularnewline
\shadeRow   ~~&   &   &   (0.95)&   (1.02)&   (1.00)&   (0.95)&   (0.92)&   (0.92)&   (0.93)&   (0.93)\tabularnewline
   ~~&   sHS&   &   1.00&   1.07&   1.11&   1.01&   0.95&   0.94&   0.96&   1.01\tabularnewline
   ~~&   &   &   (1.00)&   (1.06)&   (1.09)&   (1.02)&   (0.97)&   (0.96)&   (0.97)&   (0.97)\tabularnewline
\shadeRow   ~~&   sHS (approx.)&   &   0.95&   1.04&   0.98&   0.94&   0.92&   0.92&   0.94&   0.97\tabularnewline
\shadeRow   ~~&   &   &   (0.95)&   (1.03)&   (1.01)&   (0.96)&   (0.93)&   (0.92)&   (\textbf{0.93})&   (\textbf{0.93})\tabularnewline
\midrule
\multicolumn{11}{c}{\textbf{Nelson-Siegel TVP-VAR with the conventional \cite{primiceri2005time} setup}}\tabularnewline
~~&   &   &   1.00&   0.98&   1.02&   1.00&   0.98&   0.98&   0.99&   1.04\tabularnewline
~~&   &   &   (1.02)&   (1.07)&   (1.11)&   (1.04)&   (1.01)&   (0.99)&   (0.99)&   (1.00)\tabularnewline
\bottomrule
\caption*{\scriptsize\textit{Notes}: This table displays the three-step ahead forecast performance for non-sparsified models. We focus on seven maturities ($1$y, $3$y, $5$y, $7$y, $10$y, $15$y, and $30$y) as our target variables and use a hold-out period from $2009$:$01$ to $2019$:$12$. Point forecast performance is measured by relative root mean square errors (RMSEs), while density forecast performance (shown in parentheses) by relative continuous ranked probability scores (CRPSs). We consider two different models in terms of the dimension of the (TVP-)VARs: a large model including all $30$ maturities ($M = 30$) and a small model specified as a three factor Nelson-Siegel model ($M = 3$). For the main TVP-VARs, we consider a flexible and a RW specification of $\bm W$, each with five different global-local shrinkage priors (four dynamic and one static). These TVP-VARs are estimated with two different algorithms: our proposed approximate approach and an exact algorithm. In addition, we consider the conventional TVP-VAR setup of \cite{primiceri2005time} for the Nelson-Siegel model and a set of VARs with constant coefficients. For the VARs with constant parameters, we adopt either a Minnesota or a horseshoe (HS) shrinkage prior. As overall benchmark model we choose a large VAR with constant parameters and a Minnesota prior. The red shaded rows correspond to the actual RMSE and CRPS values of this benchmark model, while the grey shaded rows correspond to models for which we use our approximate (but non-sparsified) MCMC algorithm. The best performing specification is in bold.}
\end{longtable}}

\end{appendices}
\end{document}